\documentclass[aps,superscriptaddress,reprint]{revtex4-2}

\usepackage[utf8]{inputenc}
\usepackage{xcolor}
\usepackage{pst-node}
\usepackage{graphicx}
\usepackage{amsmath, amssymb, amsthm}
\usepackage{mathtools}
\usepackage{hyperref}
\usepackage[shortlabels]{enumitem}
\usepackage{dsfont}
\usepackage{amsthm}
\usepackage{amssymb}
\usepackage{amstext}
\usepackage{amsfonts}
\usepackage{braket}
\usepackage{nicefrac}
\usepackage{extarrows} 
\usepackage{graphicx}
\usepackage{relsize}
\usepackage{soul}
\usepackage{chemfig}
\usepackage{float}
\usepackage{mwe}
\usepackage{indentfirst}
\usepackage{xurl}

\usepackage[version=3]{mhchem}

\newcommand{\Eq}[1]{Eq.\@ \ref{#1}}

\newcommand{\br}[1]{\mathbf{#1}}

\begin{document}

\title{Orbital optimization of large active spaces via AI-accelerators} 

\author{\"Ors Legeza}
\email{legeza.ors@wigner.hu }
\affiliation{%
Strongly Correlated Systems Lend\"ulet Research Group,
Wigner Research Centre for Physics, H-1525, Budapest, Hungary
}%
\affiliation{
Institute for Advanced Study, Technical University of Munich, Germany, Lichtenbergstrasse 2a, 85748 Garching, Germany
}
\affiliation{Parmenides Stiftung, Hindenburgstr. 15, 82343, Pöcking Germany 
}

\author{Andor Menczer}
\affiliation{%
Strongly Correlated Systems Lend\"ulet Research Group,
Wigner Research Centre for Physics, H-1525, Budapest, Hungary
}%

\affiliation{%
Eötvös Loránd University, P\'azm\'any P\'eter S\'et\'any 1/C, 1117 Budapest, Hungary
}%

\author{Ádám Ganyecz}%
\affiliation{%
Strongly Correlated Systems Lend\"ulet Research Group,
Wigner Research Centre for Physics, H-1525, Budapest, Hungary
}%

\author{Mikl\'os Antal Werner}
\affiliation{%
Strongly Correlated Systems Lend\"ulet Research Group,
Wigner Research Centre for Physics, H-1525, Budapest, Hungary
}%

\author{Korn\'el Kap\'as}
\affiliation{%
Strongly Correlated Systems Lend\"ulet Research Group,
Wigner Research Centre for Physics, H-1525, Budapest, Hungary
}%

\author{Jeff Hammond}
\email{jeffpapers@nvidia.com}
\affiliation{%
NVIDIA Helsinki Oy, Porkkalankatu 1, 00180 Helsinki
}%

\author{Sotiris S. Xantheas}
\email{Sotiris.Xantheas@pnnl.gov}
\affiliation{%
Advanced Computing, Mathematics, and Data Division, Pacific Northwest National Laboratory, Richland, Washington 99354, USA}%
\affiliation{%
Department of Chemistry, University of Washington, Seattle, WA 98195, USA}

\author{Martin Ganahl}
\email{martin.ganahl@sandboxaq.com}
\affiliation{SandboxAQ, Palo Alto, California, USA}%

\author{Frank Neese}
\email{neese@kofo.mpg.de}
\affiliation{
Max-Planck Institut für Kohlenforschung, Kaiser-Wilhelm-Platz 1,\\ D-45470 Mülheim an der Ruhr, Germany}

\date{\today}% It is always \today, today,

\begin{abstract}
We present an efficient orbital optimization procedure that combines the highly GPU accelerated, spin-adapted density matrix renormalization group (DMRG) method with the complete active space self-consistent field (CAS-SCF) approach for quantum chemistry implemented in the ORCA program package. 
Leveraging the computational power of the latest generation of Nvidia GPU hardware, we perform CAS-SCF based orbital optimizations for unprecedented CAS sizes of up to 82 electrons in 82 orbitals [CAS(82,82)] in molecular systems comprising of active spaces sizes of hundreds of electrons in thousands of orbitals. 
For both the NVIDIA DGX-A100 and DGX-H100 hardware, we provide a detailed scaling and error analysis  of our DMRG-SCF approach for benchmark systems consisting of polycyclic aromatic hydrocarbons and iron-sulfur complexes of varying sizes. 
Our efforts demonstrate for the first time that highly accurate DMRG calculations at large bond dimensions are critical for obtaining reliably converged CAS-SCF energies. For the more challenging iron-sulfur benchmark systems, we furthermore find the optimized orbitals of a converged CAS-SCF calculation to depend more sensitively on the DMRG parameters
than those for the polycyclic aromatic hydrocarbons. The ability to obtain converged CAS-SCF energies and orbitals for active spaces of such large sizes within days reduces the challenges of including the appropriate orbitals into the CAS or selecting the correct minimal CAS, and may open up entirely new avenues for tackling strongly correlated molecular systems. 
\end{abstract}

\maketitle

\section{Introduction}

Since their early days more than 50 years ago, computational quantum chemistry methods have nowadays become a powerful and indispensable tool assisting scientists unraveling the mysteries of the microscopic world of atoms and molecules. Through substantial advances in algorithmic and hardware development,
today's quantum chemistry algorithms are able to accurately model molecules and materials with hundreds or thousands of electrons and orbitals, often requiring only moderate computational resources. A key element of success in this context has been the combination of groundbreaking, innovative new algorithms coupled with novel hardware advances.
Today, strongly correlated systems i.e. multi-reference systems, have become the main frontier in computational quantum chemistry. While modern quantum chemistry methods are nowadays routinely applied to compute properties of large, single-reference molecules~\cite{orca-2020,nwchem-2020,molpro-2012,Gaussian-2016,qchem-2015, ADF-2001, VASP-1996, pyscf-2020, terachem-2021, quantum_espresso-2009, quantum_espresso-2017, quantum_espresso-2020, Pederson-2023}, multi-reference quantum systems still pose significant challenges for computational methods, often to the point that a reliable description of molecular properties remains beyond reach.~\cite{Li-Manni-2018,Reiher-2009, Li-2019,Evangelista-2018}. Reliable ab-initio computational prediction of properties of such materials and molecules is considered the holy-grail of quantum chemistry, and lies at the center of contemporary research in classical and quantum approaches to computational quantum chemistry.\cite{Yudong-2019,Bauer-2020,Baiardi-2020,Li-2019}

In the context of multi-reference quantum chemistry, tensor networks methods \cite{Affleck-1987, White-1992b} are a particularly successful class of classical algorithms \cite{White-1999,Chan-2002a,Legeza-2003a} able to accurately model quantum systems with strong, local correlations\cite{
Legeza-2008,Chan-2008,Yanai-2009,Kurashige-2009,Marti-2010c,Wouters-2014a,Szalay-2015a,Chan-2016,Baiardi-2020,Cheng-2022}. Over the past decades they have emerged as the gold-standard method for strongly correlated quantum systems in low dimensions\cite{Schollwock-2005,Schollwock-2011, Verstraete-2004,Corboz-2009, Vidal-2007,Verstraete-2023}, and today often serve as benchmark tools against which novel computational approaches need to be tested\cite{Chan-2016, Baiardi-2020}. A key role is played in this context by the so-called density matrix renormalization group (DMRG) algorithm ~\cite{White-1999}, a variational optimization algorithm over the space of so-called matrix product states, a simple but very powerful tensor network.  
In this work we employ a highly GPU-efficient implementation of the DMRG algorithm as a complete active space (CAS) solver within a complete active space self-consistent field (CAS-SCF) approach~\cite{Siegbahn-1980,Roos-1980,Siegbahn-1981} 
for both closed and open shell systems.
We demonstrate that by replacing the conventional CI solver with the massively parallel hybrid CPU-GPU DMRG method~\cite{Menczer-2023b,Xiang-2024} the quality of the resulting DMRG-SCF~\cite{Zgid-2008c,Yanai-2009,Liu-2013,Wouters-2014b,Ding-2023} framework can be raised to a significantly higher level, leading to a well controlled and stable convergence of the CAS-SCF method, while simultaneously targeting active spaces of unprecedented size and complexity. This enables us to apply the CAS-SCF method to complex chemical problems far beyond the ones currently possible, completed in a fraction of the time and cost compared to alternative approaches. Moreover, the accessible large bond dimensions and active space sizes make it possible to carry out a detailed error analysis of the DMRG-SCF procedure as a function of CAS size, bond-dimension, accuracy settings of subroutines, and other tunable hyper-parameters.

We utilize a message passing interface (MPI) based massively parallel implementation of the SCF procedure via the Orca program package,~\cite{orca-2020} which is coupled efficiently with our hybrid CPU-GPU DMRG implementation.~\cite{Menczer-2023a,Menczer-2023b,Menczer-2024c} 
We focus on results obtained on a single node supplied with eight NVIDIA A100 or H100 graphics processing units leading to a factor of up to 20-70 speedup compared to a 128 CPU core single node implementation. 

Performing large-scale DMRG calculations on the optimized CAS space with even higher accuracy thresholds together with the tailored couple cluster (TCC) implementation, a significant part of the dynamic correlation can also be recovered. ~\cite{Veis-2016} The latter also support LPNO~\cite{Demel-2015} and DLPNO treatments~\cite{Brabec-2018,Lang-2019} in the ORCA software package suited for several thousands of orbitals.~\cite{Lang-2020,Antalik-2020}
Therefore, our approach paves the way for simulating strongly correlated multi-reference molecular systems with several thousands of electrons and orbitals as a routine daily task. Extensions towards multi-Node/multi-GPU DMRG implementations reaching petaflops performance is straightforward, however the details of this implementation will be discussed in a follow-up work.~\cite{Menczer-2024d}

\section{Methods}
\label{sec:theo}
In this section, we briefly overview the underlying theory of the applied DMRG-SCF method focusing on the various error sources, technical aspects related to its massive parallelization, and background on the molecular systems is was applied. 

\subsection{Complete active space SCF}
Within the Born-Oppenheimer approximation, the non-relativistic quantum chemical many-body Hamiltonian is given (in atomic units) by
\begin{align}
    H = \sum_{pq}h_{pq}\hat e_{pq}+ \sum_{pqrs} V_{pqrs} \hat e_{pqrs} \, , ~ \mathrm{with}\\
    \hat e_{pq} \equiv \sum_{\sigma} a_{p\sigma}^{\dagger}a_{q\sigma} ~~\mathrm{and}\\
    \hat e_{pqrs} \equiv \sum_{\sigma\sigma'}a_{p\sigma}^{\dagger} a_{q\sigma'}^{\dagger} a_{r\sigma'} a_{s\sigma}.
\end{align}
\\
Here, $h_{pq} \equiv (p|\hat h |q)$ and $V_{pqrs} \equiv (pq|rs)$ are the one- and two-electron integrals
\begin{align}
    (p|\hat h|q) = \int d\br{r} \phi^*_p(\br{r})\hat h(\br{r})\phi_q(\br{r})\\
    (pq|rs) = \int d\br{r}_1d\br{r}_2 \frac{\phi^*_p(\br{r}_1) \phi_q(\br{r}_1)\phi^*_r(\br{r}_2)\phi_s(\br{r}_2)}{\lVert\br{r}_1 - \br{r}_2\rVert}
\end{align}
with $\hat h(r)$ describing both the kinetic energy of the electrons as well as an external one-body potential, including the potential energy of the charged nuclei.
$\phi_p(\br{r})$, $p=1,\dots, N$, are a set of $N$ orthonormal single electron wavefunctions, 
and $a^{\dagger}_{p\sigma}$ and $a_{p\sigma}$ are fermionic creation and annihilation operators of electrons with spin $\sigma$ in orbital $p$. 

The Multi Configurtion SCF (MCSCF) method aims at incorporating static electronic correlations into a Hartree-Fock or DFT calculation both by variationally incorporating multiple Slater determinants into a variationally optimized wave function and by optimizing the single-particle orbitals $\phi_p(\br{r})$ of the Slater determinants at the same time. 
The orbitals $p=1, \dots, N$ are grouped into three distinct, non-overlapping sets of core (doubly occupied), active and virtual (unoccupied) orbitals. The active space (AS) orbitals that can have occupations of 0, 1, or 2 are chosen with the intent to capture the most prominent many-body effects of the problem at hand. Typically, the AS includes part of or the entirety of valence orbitals. Their corresponding HF single-particle energies are typically located around the Fermi level of the HF solution. In the complete active space SCF (CAS-SCF) approach, the variational optimization is carried out over all possible Slater determinants that can be obtained from redistributing electronic occupations within the AS only, while in the SCF, simultaneously,  all single-electron orbitals are also unitarily varied.
Here, we remark, that ORCA is using  configuration state functions (CSFs) not determinants but in this context, this is an insignificant detail.

Within a given basis set $\phi_p$, a general CAS wave function has the form 
\begin{align}
    \ket{\Psi} = \sum_{I \in AS}c_I\ket{\Phi_I}
\end{align}
where the sum over $I\in AS$ runs over the Slater determinants $\ket{\Phi_I}$ obtained from all possible electron configurations within the active space, and $c_I$ are the expansion coefficients of each Slater determinant. Additionally, in the CAS-SCF approach, all single-particle orbitals $\phi_p$ are unitarily optimized using a single-body rotation matrix $U \equiv e^{-\hat \kappa}$ with $\hat \kappa\equiv \sum_{p>q} \kappa_{pq} (\hat e_{pq} - \hat e_{qp})$ on top of the expansion coefficients $c_I$ in order to minimize the energy
\begin{align}
   &E = \bra{\Psi}U^{\dagger}H U \ket{\Psi} = \nonumber\\
    &\sum_{pq}h_{pq}\bra{\Psi}U^{\dagger}\hat e_{pq}U\ket{\Psi} +
    \sum_{pqrs} V_{pqrs}\bra{\Psi}U^{\dagger}\hat e_{pqrs}U\ket{\Psi}
\end{align}
from which the optimal parameters are obtained as
\begin{align}
    U_{opt}, \ket{\Psi_{opt}} = \arg\min_{U, c_I}\bra{\Psi}U^{\dagger} H U \ket{\Psi}.
\end{align}

Instead of optimizing both the CI coefficients $c_I$ and the unitary $U$ simultaneously, practical CAS-SCF implementations employ an alternating scheme of a CAS-CI step optimizing the $c_I$ coefficients at fixed $U$ followed by an orbital optimization of the unitary $U$ at fixed $c_I$.~\cite{Kollmar-2019}
In this work we employ the density matrix renormalization group (DMRG) method as the CAS-CI solver, and use ORCA to perform the orbital optimization step. The gradient of the energy with respect to the orbital rotation matrix can be obtained from the one and two electron reduced density matrices 
\begin{align}
    \tilde{\gamma}_{pq} = \bra{\Psi}\hat e_{pq} \ket{\Psi}\label{eq:onerdm}\\
    \tilde{\Gamma}_{pqrs} = \bra{\Psi}\hat e_{pqrs} \ket{\Psi}\label{eq:twordm}
\end{align}
which can be computed efficiently from the CAS-CI wavefunction obtained via DMRG.~\cite{Zgid-2008b} The iteration is stopped once the gradient for the orbital rotations is sufficiently small. 
In general, the threshold depends on the chosen convergence tightness. In the CASSCF module of the ORCA program for ``TightSCF'' it is $\times 10^{-3}$ on the gradient and at the same time $\times 10^{-7}$ on the energy change.

Regarding the superiority of CAS-SCF over CAS-CI we recall that variationally optimizing the CI coefficients and orbital rotations at the same time offers a number of advantages. First of all, derivatives of variational wavefunctions are far easier to calculate than for non-variational wavefunctions. Secondly, enforcing orbital optimization (usually) leads to a very well defined wavefunction that eliminates possible arbitrariness. Therefore, the results of a CAS-SCF calculation are typically of much better quality compared to those of a CAS-CI approach. 

An important ingredient, however,
for obtaining qualitatively and quantitatively accurate results with CAS-CI is the inclusion of the appropriate orbitals in the AS. This is far from a trivial task and usually requires the intuition of a skilled quantum chemist familiar with the intricacies of the problem at hand. We do, however, mention here recent efforts for both the automatic selection of active spaces \cite{Khedkar-2019, Stein-2016} and the development of powerful tools that assist with the selection of the initial active space orbitals (e.g. AVAS \cite{Elvira-2017}).
CAS-SCF on the other hand is computationally much more challenging but is more resilient to picking the wrong orbitals in the active space since the orbital update may mix in orbital components that have originally been left out. 
Most definitely, a ``self-repairing'' active space is rare. If one chooses a bad starting point, which is really easy, one most certainly ends up with nonsensical results. It is rather an art form to make sure that the initially picked orbitals stay in the active space. ORCA has a lot of infrastructure to try to ensure that, but the variational principle can be merciless. Ultimately, the division of orbitals into an inactive, active, and virtual space is very non-natural, and mathematics ``rewards" us for insisting on this choice with convergence problems of all kinds.

The two main bottlenecks of CAS-SCF are the computation of the CAS-CI wavefunction at fixed $U$, and the computation of one and two-body reduced density matrices $\tilde{\gamma}_{pq}$ and $\tilde{\Gamma}_{pqrs}$. The main contribution of this work is the application of our novel GPU-accelerated approach for computing the reduced density matrices, in combination with a GPU-accelerated DMRG CAS-CI solver. Leveraging the latest Nvidia supercomputing hardware, and ORCA's efficient parallelization techniques for orbital optimization, this enables us to run CAS-SCF calculations for more than 80 electrons in 80 orbitals within a day or so, far beyond any other current state-of-the-art approach. The ability to use such large active spaces
with CAS-SCF has the potential to substantially reduce the burden of researchers to pick the right orbitals for their problem, and may help 
to attack challenging strongly correlated molecular systems that are beyond the scope of conventional methods.

\subsection{Tensor network states and density matrix renormalization group}

In the following, we consider a system of $N$ Hilbert spaces $\ket{i_p}, p = 1,\dots, N$, with $i_p = 0,1,2,3$ corresponding to states with no electron, one down electron, one up electron, and one up and one down electron, respectively. In the context of this manuscript, the states $\ket{i_p}$ correspond to the single-electron orbitals $\phi_p$ obtained from a Hartree-Fock or DFT calculation. 
For example, a restricted Hartree-Fock wave function is a simple product state obtained by filling all orbitals $\ket{i_p}$ with low (HF orbital) energies $\epsilon_p, p<p^*$ with two electrons, and keeping all other orbitals empty: 
\begin{equation}
    \ket{\Psi} = \ket{3}_i \dots \ket{3}_{p^*}\ket{0}_{p^{*}+1}\dots\ket{0}_N.
\end{equation}
For a given set of inactive but occupied orbitals $\ket{i_{o_k}, k = 1,\dots,N_{IA}}$,
\begin{equation}
\ket{\Psi} = \left(\bigotimes_{k=1}^{N_{IA}}\ket{3}_{o_k}\right)\sum_{\{i_{p_k}\}} \psi_{i_{p_1}\dots i_{p_{N_A}}} \ket{i_{p_1}\dots i_{p_{N_A}}}\left(\bigotimes_{k=1}^{N_{UO}}\ket{0}_{q_k}\right).
\end{equation}
For the sake of brevity, we will relabel all orbitals such that the first $N_A$ orbitals $\ket{i_k}, k=1,\dots,M_A$ correspond to the active space orbitals, 
and omit the explicit reference to the trivial tensor products over inactive and empty spaces, i.e.
\begin{equation}
  \ket{\Psi} =\sum_{\{i_{1} \dots i_{N_A}\}} \psi_{i_1\dots i_{N_A}} \ket{i_1\dots i_{N_A}}.\label{eq:psi}
\end{equation}
Matrix product state (MPS) wave functions are a special class of many body wave functions parameterized by a set of $N_A$ 
tensors $A_{\alpha_{p-1}\alpha_{p}}^{i_p}$ of shape $(4, D_{p-1}, D_p)$, such that the many body wave function \Eq{eq:psi} takes on the special form
\begin{equation}
  \psi_{i_1\dots i_{N_A}} = \sum_{\{\alpha_p\}}[A_1]_{1\alpha_1}^{i_1} [A_2]_{\alpha_1\alpha_2}^{i_2} \dots [A_{N_A}]_{\alpha_{N_A-1}1}^{i_{N_A}}.
\end{equation}
For a fixed set of orbital indices $i_1,\dots,i_{N_A}$, the amplitude $\psi_{i_1\dots i_{N_A}}$ reduces to a sequence of matrix products, giving the ansatz its name~\cite{Schollwock-2011}. 
The largest of the tensor dimension $ D \equiv \max_k D_k$ is called the bond dimension of the MPS. The expressiveness of the MPS wave function
increases with increasing bond dimension $D$. MPS wave functions with small values of $D$ are particularly well suited to represent wave functions with low
entanglement. Note that any state \Eq{eq:psi} can be cast into an MPS form, though the bond dimension required 
for a faithful representation might be exponentially large in $N_A$.
In this work we employ the density matrix renormalization group (DMRG) method as a CAS-CI solver~\cite{White-1992b,White-1999}. In its modern formulation, the DMRG can be understood as a variational
minimization over the space of MPS~\cite{Ostlund-1995}. Specifically, we use it to variationally determine an approximation of the ground state of the CAS-CI Hamiltonian 
\begin{equation}
  E_{opt} = \min_{\ket{\Psi}}\frac{\bra{\Psi}H\ket{\Psi}}{\braket{\Psi|\Psi}}.
\end{equation}
From the optimized MPS wave function $\ket{\Psi_{opt}}$, the one- and two-electron reduced density matrices \Eq{eq:onerdm} and \Eq{eq:twordm} can be efficiently
computed using standard techniques \cite{Zgid-2008b}. One of our main contributions in this work is the utilization of our GPU accelerated method to compute
$\tilde{\gamma}_{pq}$ and $\tilde{\Gamma}_{pqrs}$ with substantial speed ups compared to existing implementations, which we put to use for running CAS-SCF calculations at large bond dimensions to obtain stable convergence with high accuracy, and at unprecedented AS sizes of up to 82 electrons in 82 orbitals in about a day or so.

\subsection{Conservation of the total spin in the DMRG-SCF procedure}
\label{sec:na}

Our hybrid CPU-multiGPU DMRG code can handle both Abelian and non-Abelian quantum numbers~\cite{Menczer-2023a,Menczer-2023b}, thus a given state with total spin can also be targeted via the DMRG-SCF protocol. In this case, the DMRG bond dimension $D$ refers to the number of multiplets kept to represent the left or the right DMRG block
\cite{Mcculloch-2002, Toth-2008, Sharma-2012a, Keller-2016,Gunst-2019, Werner-2020su3}. This in general, leads to a factor of two to three reduction in $D$ for a singlet ($S_\mathrm{tot}=0$) state to reach similar accuracy as obtained with a strict U(1) implementation, while for states with higher total spin, this reduction can be even larger.
This reduction provides a considerable speedup in the DMRG part of the algorithm. In addition, since no spin contamination can happen, the orbital optimization procedure becomes more stable and robust. In our implementation, the SU(2) related Clebsch-Gordan layer~\cite{Werner-2020su3} is fully decoupled from the MPS layer, thus there is no overhead in the matrix and tensor algebra when massive parallelization is utilized~\cite{Menczer-2023b,Menczer-2024a}. In the rest of the paper, all bond dimensions $D$ are reported as SU(2) multiplets, with the corresponding U(1) bond dimensions, ${\tilde D}$, indicated separately where applicable.

As our non-Abelian DMRG implementation treats the Hilbert-space at the multiplet level, the symmetry-coupled form of the one and two electron reduced density matrices are determined, that are built based on the spin-$\frac{1}{2}$ Wigner-Eckart tensor operators formed from creation and annihilation operators:
\begin{equation}
    f_p^\dag = (a_{p\uparrow}^\dag, a_{p\downarrow}^\dag)  , \quad  f_p = (a_{p\downarrow}, -a_{p\uparrow} ) \; .
\end{equation}
The spin-coupled form of the one electron reduced density matrix is simply the expectation value of the spin-0 combination of the $f_p^\dag f_q$ product, 
\begin{eqnarray}
    \gamma_{pq} & & = \bra{\Psi} [f^\dag_p f_q]_0 \ket{\Psi} =  \nonumber \\
    & & \bra{\Psi} \frac{1}{\sqrt{2}} \left(f_{p,1}^\dag f_{q,2} - f_{p,2}^\dag f_{q,1} \right) \ket{\Psi} = -\frac{1}{\sqrt{2}} \tilde{\gamma}_{pq} \label{eq:rdm1_su2}
\end{eqnarray}
Here and in the following $[AB]_J$ denotes the spin-$J$ tensor operator formed from the product of tensor operators $A$ and $B$, which is calculated as the linear combination of simple operator products multiplied by the appropriate Clebsch-Gordan coefficients like in Eq.~\eqref{eq:rdm1_su2}. In case of the two electron reduced density matrix, two channels are formed depending on the total spin created by the $f^\dag_p f^\dag_q$ product,
\begin{eqnarray}
    \Gamma_{pqrs}^{(0)} &=& \bra{\Psi} \left[[f^\dag_p f^\dag_q]_0 [f_r f_s]_0 \right]_0 \ket{\Psi} \nonumber \\
    \Gamma_{pqrs}^{(1)} &=& \bra{\Psi} \left[[f^\dag_p f^\dag_q]_1 [f_r f_s]_1 \right]_0 \ket{\Psi}  \; ,
\end{eqnarray}
and from these the original \eqref{eq:twordm} form of the density matrix is simply
\begin{equation}
   \tilde{\Gamma}_{pqrs} = 2 \sqrt{3} \Gamma^{(1)}_{pqrs} - \Gamma^{(0)}_{pqrs} \; .
\end{equation}
Having the $\tilde{\gamma}_{pq}$ and $\tilde{\Gamma}_{pqrs}$ density matrices at hand, CASSCF optimization of ORCA can be called in the same way as in the U(1) symmetric case.

\subsection{Error sources}
\label{sec:error}

In the DMRG-SCF method there are various error sources that can accumulate during the procedure and these can even counteract with each other leading to unstable convergence. The CASSCF theory is based on the full-CI, i.e., the exact solution of the CAS, thus convergence is controlled and determined solely by the parameters of the applied gradient descend methods. In the ORCA program package, there are several options to achieve fastest convergence~\cite{orca-2020}. 
For the CASSCF procedure the default ``TightSCF'' setting
corresponds to $\varepsilon_{\rm ECT} = \times10^{-7}$, and $\varepsilon_{\rm OGC} = \times10^{-3}$.

In contrast to these, the DMRG-SCF procedure brings in another error source since the CAS solution is only approximate and determined mainly by the DMRG bond dimension, $D$ ~\cite{Legeza-1996,Legeza-2003a,Schollwock-2005,Noack-2005}. For the ab initio DMRG variant, there are also further factors that influence
the convergence, thus various algorithmic solutions developed in the past two decades~\cite{Kurashige-2009,Szalay-2015b,Baiardi-2020,Cheng-2022} must be utilized to reach the a priori set error margin with fastest convergence rate. Only if the error of the DMRG solution is kept below the error settings of the SCF procedure could the DMRG-SCF procedure lead to stable convergence. This, however, usually requires large $D$ values making the DMRG-SCF procedure impractically too long when large CAS spaces are considered without massive parallelization 
as has been faced in most of the previous attempts.

In addition, when only an approximate CAS solution is used instead of the full-CI wave function, then technically active--active rotations should also be utilized. This, however, is usually neglected as they lead to numerical instabilities as the full-CI limit is approached.
A way to circumvent this problem is to utilize more advanced DMRG protocols based on fermionic mode optimization~\cite{Krumnow-2016,Krumnow-2021,Mate-2022,Petrov-2023,Werner-2025} which also optimizes the underlying one particle basis in the CAS. 
This method also reduces the level of entanglement encoded in the CAS wave function, thus same accuracy can be achieved with significantly lower DMRG bond dimension. In addition, it is very stable and robust and even stationary conditions can be fulfilled~\cite{Friesecke-2023}. 
After convergence is reached the resulting one- and two-particle reduced density matrices can be back rotated to the original CAS basis~\cite{Krumnow-2021,Menczer-2024a} and passed to the SCF solver. Here, however, we do not employ such mode optimization, in order to keep the procedure simple.

\subsection{Parallelization and technical aspects}
\label{sec:parallel}

The wall time of the DMRG-SCF procedure comes from two main components, i.e., the total time spent on the SCF optimization procedure and on the solution of the selected CAS via the DMRG method. Since both scale with system size as $N_A^4$ where $N_A$ stands for the number of orbitals %
a massive parallelization is mandatory to keep computational time feasible.

On the one hand, the massively parallel implementation of the SCF module based on message passing interface (MPI) in the ORCA program package can scale up to twenty-forty CPU cores easily~\cite{orca-2020}. 
On the other hand, a massively parallel hybrid CPU-multiGPU DMRG method~\cite{Menczer-2023a,Menczer-2023b}, introduced by some of us recently, can reach even a quarter petaflops on a single node by utilizing NVIDIA AI accelerators~\cite{Menczer-2024b}. Consequently, the cubic scaling of the wall time with bond dimension in the DMRG algorithm can be reduced to linear scaling by employing parallelization over increasing number of GPUs for a broad range of $D$ values~\cite{Menczer-2023a,Menczer-2023b}. 
Similarly, the calculation of the one- and two-particle reduced density matrices has been adapted to multi-GPU systems ~\cite{Menczer-2024e}, showing almost perfect scaling behavior with increasing system size, DMRG bond dimension and number of GPUs.
This provides us with the ability to use much larger values of bond dimension $D$ than previously achieved, at a marginal increase in computation time. The opportunity to use such large $D$ values has two key advantages. First, the number of DMRG sweeps to achieve a desired accuracy can be reduced significantly. Since the bulk of the runtime of a DMRG-SCF iteration step is spent in the DMRG-CI solver, this translates into a major speedup for the DMRG-SCF procedure itself.
Second, we empirically find that the higher accuracy resulting from large bond dimension calculations significantly improves the convergence behavior of the SCF procedure, resulting in a drastic reduction of required number of SCF iteration steps to achieve convergence. These two key properties allow us to obtain converged DMRG-SCF calculations for CAS sizes far beyond what has so far been reported in the literature, within run times ranging from minutes to $\approx 30$ hours.

To utilize all benefits of parallelization we perform calculations on 
a dual AMD EPYC 7702 CPUs with $2\times 64$ cores combined with eight NVIDIA A100-SXM4-40GB devices and on 
a dual Intel Xeon Platinum 8481C CPUs with  $2\times 52$ cores combined with eight NVIDIA H100-HBM3-80GB devices.
For the ORCA suite 24-48 MPI processes are allocated, depending on the estimated memory requirement of a single process in the SCF module, while for the DMRG part all available number of threads and GPUs are utilized. 

\subsection{Numerical procedure}
\label{sec:numproc}

Our DMRG-SCF method relies on our hybrid CPU-GPU DMRG implementation interfaced to the ORCA program package. For a given CAS we first perform a few DMRG optimization steps with low bond dimension, $D=\{16,32,64\}$
to obtain an optimal orbital ordering~\cite{Legeza-2003c,Barcza-2011}. Next, large scale DMRG-SCF is applied by using either fixed $D\in\{128,\ldots,4096\}$, or fixed truncation error $\varepsilon_{\rm TR}$ using the dynamic block state selection (DBSS) approach ~\cite{Legeza-2003a}. Note that for reliable convergence of the SCF procedure, $\varepsilon_{\rm TR}$ should be at least an order of magnitude smaller than ORCA's $\varepsilon_{\rm OGC}$. 
The DMRG procedure is terminated when the energy difference between three subsequent sweeps falls below an a-priory set error margin of $\varepsilon_{\rm sweep}=10^{-3}$.
The residual error of the L\'anczos and Davidson diagonalization methods has been set to $10^{-7}$.
When only U(1) symmetries are enforced, we use the dynamically extended active space (DEAS) procedure ~\cite{Legeza-2003c} to speed up convergence. 
For SU(2) symmetries, the DEAS method is currently being implemented and is expected to lead to a significant reduction in the required number of sweeps, and hence an additional speedup of the DMRG-SCF approach.

In the next section we present results obtained both at fixed bond dimension, as well as  at fixed truncation error $\varepsilon_{\rm TR}=10^{-4}$, using the DBSS method.
The number of DMRG sweeps needed to reach convergence typically varies between ten and thirty.
SU(2)$\times$ U(1) symmetry is enforced exactly using the spin-adapted DMRG approach, and one- and two-particle reduced density matrices are converted to the U(1)$\times$ U(1) representation before passing them to the ORCA program package as discussed in Sec.~\ref{sec:na}.
For sake of completeness, we also compare results obtained via the non-spin adapted workflow for the Heptacene test system.

\section{Preparation of benchmark systems}
\label{sec:systems}
\begin{figure}
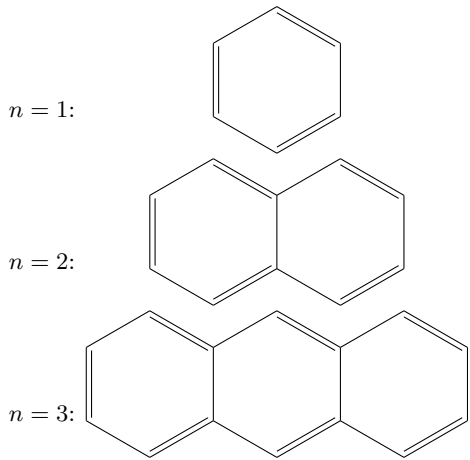

    \centering
    \def\arraystretch{5}
\begin{tabular}{lc}
     $n=1$: &\chemfig{*6(-=-=-=)}\\
     $n=2$: &\chemfig{*6(-=*6(-=-=-)-=-=)}\\
     $n=3$: &\chemfig{*6(-=*6(-=*6(-=-=-)-=-)-=-=)}
\end{tabular}
    \caption{Smallest molecules of the \ce{PAH} series for ring numbers $n=1,2$, and $3$}
    \label{fig:pah}
\end{figure}

As a first group of test systems we focus on polycyclic aromatic hydrocarbons (PAHs), which have been the subject of extensive previous numerical studies~\cite{Hammond-2007,Hajgato-2009,Zade-2010,Pelzer-2011,Aiga-2012,Rivero-2013,Plasser-2013}. Specifically, we use linear chains of aromatic rings (see Fig. \ref{fig:pah}; benzene, naphthalene, anthracene, tetracene, etc., C$_{4n+2}$H$_{2n+4}$) of up to 20 rings. The geometries were optimized with the B3LYP functional~\cite{Becke-1992,Becke-1993} and cc-pVDZ basis set~\cite{Dunning-1989,Woon-1993}. From the corresponding orbitals, a subset containing 4$n$+2 orbitals was chosen manually, with $n$ the number of elementary building blocks (rings). The subset was chosen such as to represent the delocalized $\pi$-system over the entire PAH. The selected orbitals formed the initial active space for subsequent DMRG-SCF calculations. PAH molecules are typically single-reference and hence serve as ideal benchmark systems for method development.

The second group of test systems contains iron-sulfur clusters, which are found in various proteins like ferredoxins, as well as various hydrogenases, dehydrogenases, reductases, and nitrogenases~\cite{Lippard-1994}. 
The simplest iron-sulfur system is the 2Fe-2S cluster, which has a planar \ce{Fe2S2} core and is usually coordinated by four ligands leading to a tetrahedral structure around the iron atoms. 

\begin{figure}
%    \centering
\def\arraystretch{5}
\begin{tabular}{lc}
  $n=1$:   & \chemfig{HS-[1]Fe(-[3]HS)(<:[0.5]SH)(<[-0.5]SH)} \\
  $n=2$:   & \chemfig{HS-[1]Fe(-[3]HS)(<:[0.5]S)<[-0.5]S>[0.5]Fe(<:[3.5]S)(-[1]SH)-[-1]SH} \\
  $n=3$:   & \chemfig{HS-[1]Fe(-[3]HS)(<:[0.5]S)<[-0.5]S>[0.5]Fe(<:[3.5]S)(-[1]S)-[-1]S-[1]Fe(-[3]S)(<:[0.5]SH)(<[-0.5]SH)}
\end{tabular}

    \caption{\ce{Fe2S2} cluster series for $n=1,2$, and $3$}
    \label{fig:fe2s2}
\end{figure}

A scalable test system can be made by repeating the \ce{Fe2S2} motif and closing with -SH groups. We created the series of Fe$_n$S$_{2n-2}$(SH)$_4^{n-}$ clusters with $1\leq n\leq 3$. The first three elements are shown in Fig. \ref{fig:fe2s2}. The first species, Fe(SH)$_4^-$, corresponding to $n=1$ does not contain the \ce{Fe2S2} structure.
For $n=2$ we have the \ce{Fe2S2} motif ending with SH ligands. For $n>2$ the \ce{Fe2S2} blocks follow each other while the plane is perpendicular to the previous one.
In the oxidized state, the Fe ions are in the oxidation state of $+3$. In a tetrahedral coordination environment, the local spins located in the iron d-based molecular orbitals are all aligned parallel, leading to local S(Fe)=5/2 fragments. The different local fragments will couple antiferromagnetically to lead to overall low spin states with total spin zero, e.g. $S$=0 for the dimeric species. However, it should clearly be recognized that this is not a closed shell state but an antiferromagnetic singlet with ten unpaired electrons. Provided that localized orbitals are employed that are properly aligned to belong to the iron ions on the right- and left-hand side, a single configuration state function (CSF) dominates the CAS(10,10) wavefunction.~\cite{Izsak-2023} This single CSF has a weight of 90 percent. However, in terms of individual Slater determinants the compactness of this representation is lost and there are 252 Slater determinants belonging to the leading CSF that each show a small CI coefficient on the order of 0.02-0.03.  
In this work, we demonstrate the power of the DMRG implementation by going to much larger active spaces than the minimal CAS(10,10) that was described above. 

Geometries in the \ce{Fe2S2} series were also optimized with B3LYP functional\cite{Becke-1992,Becke-1993} and cc-pVDZ basis set\cite{Dunning-1989,Woon-1993}. Orbitals were selected from quasi-restricted orbitals (QRO). 
Three different active spaces were defined, labeled as `A', `B', and `C', and their composition is presented in Table \ref{tab:fe2s2_cas}.
`A' contains the 3d, 4s and 4p orbitals of iron, and the 3p orbitals of the sulfur atoms, however the p orbitals of sulfurs in the terminal SH groups which make the S-H bonds are excluded. 
`B' contains also the 4d orbitals of Fe, while `C' includes 3s and 3p orbitals of Fe and 3s orbitals of S-s.

\begin{table}[t]
  \centering
\begin{tabular}{lcc|ccc}
\hline
 \hline
species & orbital type & active space &number & electron & orbital \\ 
Fe & 3s & C & $n$ & 2 & 1\\
Fe & 3p & C & $n$ & 6 & 3\\
Fe & 3d & A,B,C & $n$ & 6 & 5\\ %iron: 4s2 3d6; the 3+1(from the charge) electron helps to form the Fe-S bonds.
Fe & 4s & A,B,C & $n$ & 2 & 1\\
Fe & 4p & A,B,C & $n$ & 0 & 3\\
Fe & 4d & B,C & $n$ & 0 & 5\\
S (bridging) & 3s & C &$2n-2$ & 2 & 1 \\
S (bridging) & 3p & A,B,C &$2n-2$ & 4 & 3 \\
S (terminal) & 3s & C &$4$ & 2 & 1 \\
S (terminal) & 3p & A,B,C &$4$ & 3 & 2 \\
electron   & -  & A,B,C & $n$ & 1 & 0 \\
 \hline
 \hline
\end{tabular}
\caption{Breakdown of the active space of the Fe$_n$S$_{2n-2}$(SH)$_4^{n-}$ series. For the sake of completeness, the electron leading to the correct charge is also included in this list.}
\label{tab:fe2s2_cas}
\end{table}

\section{Numerical results}

In this section, we present results obtained by various DMRG settings for polycyclic aromatic hydrocarbons and for iron-sulfur clusters. For better readability at certain parts of our analysis we label bond dimension used in the DMRG-SCF procedure with $D_{\rm opt}$ and in DMRG-CI with $D$ while in the figures such different labeling is not applied.

\subsection{Polycyclic aromatic hydrocarbons (PAHs)}
\label{sec:numres}

\subsubsection{Convergence analysis}
\begin{figure}
    \centering
    \includegraphics[width=0.48\textwidth]
    {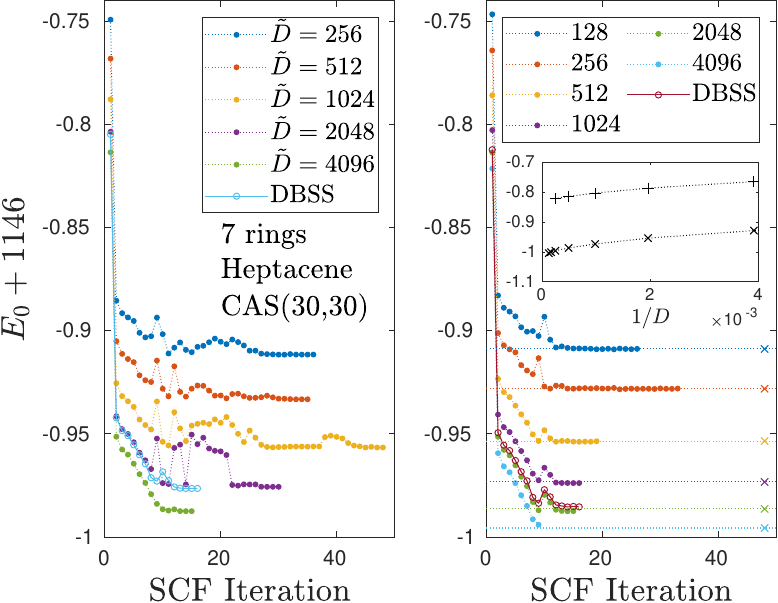}
\caption{(left) Convergence of ground state energy, $E_0$, shifted by 1146 for the heptacene on a CAS(30,30) as a function of the DMRG-SCF macro iterations for various fixed U(1) ${\tilde D_{\rm opt}}$ values and (right) via the spin adapted workflow for various fixed SU(2) bond dimensions, $D_{\rm opt}$.
Data points shown by open circle symbols were obtained via the DBSS procedure by setting ${\tilde D}_{\rm opt,min}=2048$ and $D_{\rm opt,min}=1024$, $\varepsilon_{\rm TR}=10^{-4}$ and maximum number of DMRG sweeps to ten. In the SCF module we set $\varepsilon_{\rm ECT}=10^{-7}$ and $\varepsilon_{\rm OGC}=10^{-3}$.
Symbols X together with horizontal dotted lines stand for post DMRG-SCF energy values obtained for various $D$ values using the basis optimized with $D_{\rm opt}=128$.
The inset shows the $1/D$ scaling of the shifted ground state energy, using the original non-optimized and the optimized basis with $D_{\rm opt}=128$, i.e. $E_0(D)$ and $E_0(D_{\rm opt},D)$, respectively. 
}
    \label{fig:heptacene_E}
\end{figure}

In Fig.~\ref{fig:heptacene_E} we show the convergence of the ground state energy $E_0$, (offset by 1146 for convenience of plotting) of heptacene (\ce{C30H18}) as a function of the SCF iteration steps using the ab-initio DMRG as a CAS CI-solver. 
The left panel shows results obtained using U(1) symmetry only, results in the right panel were obtained using a fully SU(2) symmetric simulation. We show results for increasing U(1) and SU(2) bond dimensions ${\tilde D}$ and $D$. 

As expected, increasing values of ${\tilde D}$ result in decreasing values for the energy in both cases. A key observation from the plots is a significant improvement of the SCF convergence with increasing DMRG accuracy (i.e. increasing bond dimension). For low values of $\tilde D=256$ (left plot) and $D=128$ (right plot) we observe non-monotonic behavior of the energy as a function of the SCF iteration step, i.e. the energy between to subsequent SCF steps can increase. This behavior is particularly pronounced for small bond dimensions and during early SCF steps, but can appear as well at later steps and for moderately large bond dimensions (see for example $\tilde D=1024$ (left panel)). For large bond dimensions, we observe fast and smooth convergence of the SCF procedure, with significantly reduced number of SCF iterations required to obtain a desired convergence level.
We note that in addition, large values of ${\tilde D}$ typically also lead to fewer DMRG-sweeps required to obtain convergence within the DMRG-CI solver.
Consequently, we empirically find a computationally optimal solution regarding accuracy and wall time by setting the truncation error margin $\varepsilon_{\rm TR}=10^{-4}$ to be an order of magnitude smaller than ORCA's $\varepsilon_{\rm OGC}=10^{-3}$, and employing the DBSS method with a minimal bond dimension $\tilde D_{\rm min}=2048$ in the DMRG-CI solver.

Enforcing SU(2) spin symmetry leads to further improvements of the convergence behavior of the SCF procedure.
Here we note that even a larger fixed $D=1024$ calculation can be stuck at a local minimum and several SCF steps were mandatory to reach final convergence.
In contrast to this, using the DBSS procedure with $D_{\rm opt,min}=1024$ results in a stable and fast convergence (open circles in Fig.~\ref{fig:heptacene_E}, right panel).

\subsubsection{Quality of the optimized basis}
An important question in the context of DMRG-SCF is in howfar the final one-particle basis at the end of the DMRG-SCF procedure depends on the bond dimension used in the DMRG-CI solver during the SCF iterations. The quality and robustness of the single-particle basis obtained at a low bond dimension can be checked by first converging the DMRG-SCF at low bond dimension, obtaining an energy $E_0(D_{\rm opt}=128)$, and then performing standard DMRG calculations with the converged basis at increasing bond dimensions, and comparing the resulting energies $E_0(D,D_{\rm opt}=128)$ with DMRG-SCF energies obtained using the same bond dimensions $D_{\rm{opt}}'=D$. Matching energies indicate that dynamic correlations included via the SCF procedure are captured well already at low-bond-dimensions. On the other hand, 
a basis error mismatch $\delta_E(D, D_{\rm opt})\equiv E_0(D,D_{\rm opt}) - E_0(D_{\rm opt}'=D)\neq 0$ indicates missing dynamic correlations at low bond-dimension DMRG-SCF calculations.
The results of these tests are shown in the right panel of Fig.~\ref{fig:heptacene_E}.
The horizontal dashed lines show energies obtained by first converging DMRG-SCF at $D_{\rm{opt}}=128$, and then increasing the bond dimension at fixed basis to $D=256, 512, 1024, 2048$ and $4096$.
For heptacene we obtain a basis error mismatch of  $\delta_E(D,D_{\rm opt})\leq 1.2\times 10^{-3}$ for all values of $D$, well below chemical accuracy of 1.6mHartree. It indicates that for single-reference systems, low bond dimension DMRG-SCF calculations already capture the bulk of dynamic correlations, in alignment with the expectation that single reference systems are typically easier to handle numerically. 
For strongly correlated systems on the other hand we expect to observe an increasing basis error with increasing bond dimension $D$. We expect that a systematic analysis of $\delta_E(D,D_{\rm opt})$ could provide a useful tool for analyzing convergence of DMRG-SCF simulations in these cases. 

In the inset of Fig.~\ref{fig:heptacene_E}(right) we show a $1/D$ scaling analysis of the (shifted) DMRG energies  using the optimized basis, $E_0(D,D_{\rm opt})$, with $D_{\rm opt}=128$
(x-symbols) and the (shifted) DMRG energies, $E_0(D)$,  obtained without the SCF procedure ({\large{+}}-symbols). 
A big difference in energy in the order of 200 milliHartree remains for all $D$ values, highlighting the importance of the SCF procedure. Using a second order polynomial fit, the $D\rightarrow\infty$ truncation-free solution can be approximately obtained.

\begin{figure}
    \centering    \includegraphics[width=0.5\textwidth]
    {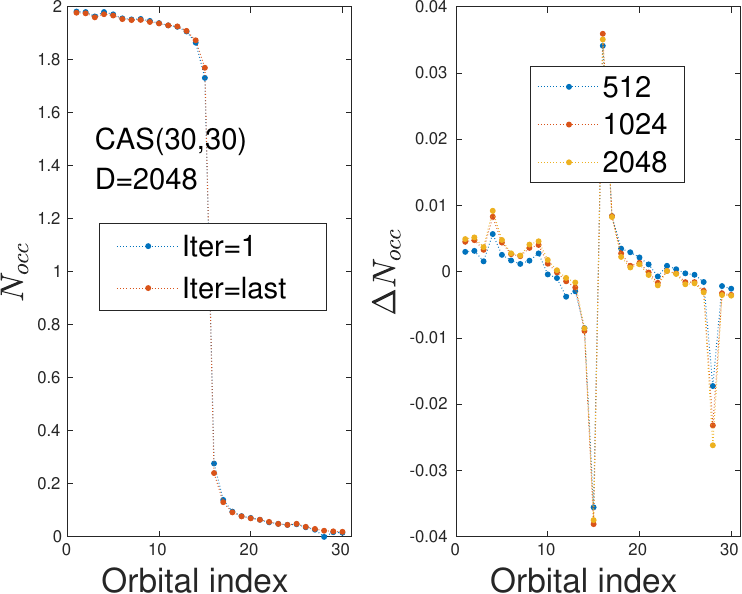}
\caption{(left) Occupation number, $N_{\rm occ}$, obtained by the first and last DMRG-SCF macro iteration for $D_{\rm opt}=2048$ and (right) the difference between the obtained $N_{\rm occ}$ profile via the first and last DMRG-SCF macro iteration for various $D_{\rm opt}$ values. In the SCF module the convergence on energy was set to $10^{-7}$ and the gradient to $10^{-3}$.}
    \label{fig:heptacene_N}
\end{figure}

The good quality of the low-$D_{\rm opt}$ optimized basis can be further understood by monitoring the evolution of the orbital occupation number, $N_{\rm occ}$, via the DMRG-SCF procedure.
In Fig.~\ref{fig:heptacene_N}(left), the occupation number is shown for the initial (first) SCF step and for the last one. This can show in general how multireference nature of a given system is changed towards
a single reference one by pushing orbital occupation towards two and zero in the vicinity of the Fermi surface. For better monitoring, i.e, to see how changes in the profile depend on the bond dimension, in
Fig.~\ref{fig:heptacene_N}(right) we present the difference between the obtained $N_{\rm occ}$ profile via the first and last DMRG-SCF macro iteration for various $D_{\rm opt}$ values. Here the systematic improvement is evident which gets less and less pronounced with increasing $D_{\rm opt}$ values. 

\subsubsection{Large active spaces}
\begin{figure}
    \centering
    \includegraphics[width=0.5\textwidth]{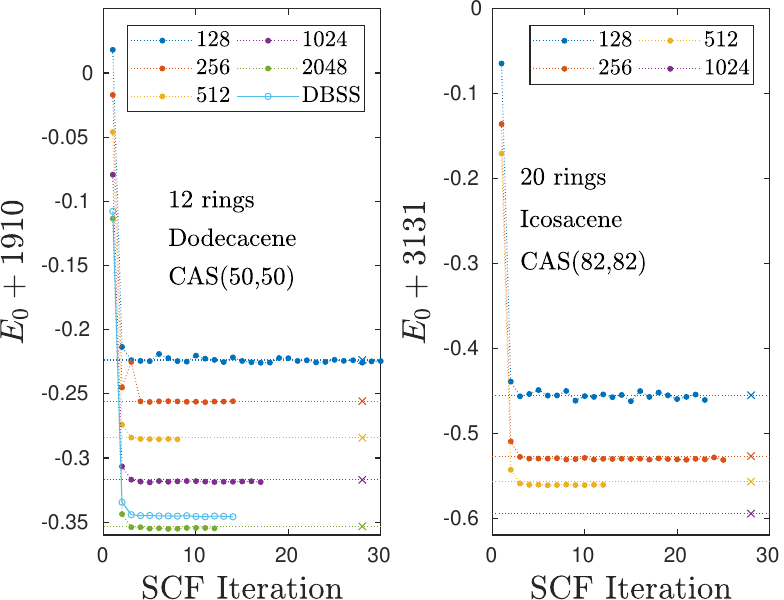}
\caption{(left panel) Similar to Fig.~\ref{fig:heptacene_E}(right), but for the Dodecacene on a CAS(50,50) and (right panel) for the Icosacene on a CAS(82,82).}
    \label{fig:dodecatene_E}
\end{figure}

In Figs.~\ref{fig:dodecatene_E} we show similar analyses to the previous section but for larger active spaces, using dodecacene(\ce{C50H28}) on a CAS(50,50) and the icosacene (\ce{C82H44}) on a CAS(82,82). Again we observe a significant reduction in SCF iteration number and a more stable convergence with increasing bond dimension $D_{\rm opt}$.
We repeat the analysis of the previous section of the basis error mismatch by performing DMRG calculations at increasing bond dimensions $D$ using the bases obtained from DMRG-SCF at $D_{\rm opt} = 128$ for both molecules. We obtain a basis error mismatch in Dodecacene
of $\delta_E(D,D_{\rm opt})\simeq1.4\times10^{-3}$ for $D=2048$ and in Icosacene $\simeq4.1\times10^{-3}$ for $D=512$ already.  
Therefore, for the larger active space CAS(82,82) the error gets more pronounced.  
This highlights the importance of basis optimization with high accuracy
for increasing CAS sizes.

From a technical point of view, it is also evident from Figs.~\ref{fig:heptacene_E} and \ref{fig:dodecatene_E} that the overall accuracy in leading order is determined by the DMRG bond dimension, i.e., by the accuracy of the CAS wave function. 
Therefore, an optimal protocol to set error margins would rely on the DBSS procedure. First, a DMRG-SCF procedure would be performed starting with a higher truncation error $\varepsilon_{\rm tr}\simeq10^{-3}$ and by setting $\varepsilon_{\rm OGC}$ to an order of magnitude larger and setting the accuracy of the residual error of the diagonalization algorithm (e.g. Davidson or L\'anczos) to an order of magnitude smaller. After convergence is reached, the DMRG-SCF procedure could be repeated by lowering $\varepsilon_{\rm tr}$ and adjusting $\varepsilon_{\rm OGC}$ and the residual error of the diagonalization algorithm accordingly.
An extrapolation, as a function of $\varepsilon_{\rm tr}$ can be performed towards the truncation-free solution just like in conventional DMRG~\cite{Legeza-1996,Legeza-2003a}. 

To provide further insights, in Table~\ref{tab:exetime} we summarize the total computational time, $t_{\rm A100}$ and $t_{\rm H100}$ obtained on DGX-A100 and DXG-H100 single nodes, respectively, via fixed $D$ and via the DBSS procedure for various parameter sets.
\begin{table}[t]
  \centering
\begin{tabular}{l|c|c|c|c|c|c|c|r}
\hline
 \hline
$\tilde N_{\rm e}$ & $\tilde N$ & CAS & $D_{\rm min}$ & $D_{\rm max}$ & ${\tilde D}_{\rm max}$ & $t_{\rm A100}$ & $t_{\rm H100}$ & Energy\\

 \hline
 198 & 510 & (30,30) & 512  & 512  & 1606  & 5.0 h  & 3.0 h& -1146.9536 \\
 198 & 510 & (30,30) & 1024 & 1024 & 3453  & 7.8 h & 5.1 h& -1146.9738 \\
 198 & 510 & (30,30) & 2048 & 2048 & 5262  & 19.7 h & 13.1 h& -1146.9875\\
 198 & 510 & (30,30) & 1024 & 5363 & 17022 & 15.7 h & 10.3 h& -1146.9853 \\
 \hline
 328 & 840 & (50,50) & 512 & 512 & 1622 & 6.5 h& 4.1 h& -1910.2853 \\
 328 & 840 & (50,50) & 1024 & 1024 & 2561 & 24.8 h & 16.3 h & -1910.3187 \\
 328 & 840 & (50,50) & 1024 & 4096 & 12826 & 63.7 h & 25.4 h& -1910.3458 \\ 
 \hline
 536 & 1368 & (82,82) & 256 & 256 & 585 & 49.4 h & 22.4 h & -3131.5315 \\
 536 & 1368 & (82,82) & 512 & 512 & 1115 & 54.6 h & 32.1 h & -3131.5613 \\  
 \hline
 \hline
\end{tabular}
\caption{Total number of electrons, $\tilde N_{\rm e}$, and orbitals, $\tilde N$, of the full orbital space, size of selected CAS, 
minimum DMRG bond dimension, $D_{\rm min}$, 
maximum DMRG bond dimension, $D_{\rm max}$, 
corresponding maximum DMRG U(1) bond dimension, ${\tilde D}_{\rm max}$,
and total time in hours for the eight A100 and H100 GPU accelerated DMRG-SCF, $t_{\rm A100}$, $t_{\rm H100}$, respectively, for the Heptacene, Dodecacene, and Icosacene obtained with $\varepsilon_{\rm TR}=10^{-4}$.
The maximum number of DMRG sweeps was set to ten. For fixed $D$ calculations
$D_{\rm min} = D_{\rm max}$.
}
\label{tab:exetime}
\end{table}
We note that our implementation shows high GPU utilization already at intermediate levels of bond dimension $D$ ~\cite{Menczer-2024b}, hence providing substantial speedups for DMRG-SCF runs already for smaller-scale problems, compared to previous implementations.

\subsection{Iron-sulfur clusters}

Next, we consider a more challenging system, built from iron-sulfur clusters (see Fig.~\ref{fig:fe2s2}), where the appearance of metal atoms requires a more elaborate numerical treatment. 
The ground state is an open shell state with spin $S=5/2$ or $S=0$ for even and odd values of $n$, respectively. The more challenging nature of these systems requires careful adjustment of the DMRG-SCF hyperparameters, and typically between 20-30 sweeps per DMRG run were necessary to achieve stable convergence in the DMRG-SCF runs.
Moreover, for sextet ground states ($S=5/2$, even values of $n$), the high multiplicity leads to a significant (roughly six times) difference between the SU(2) $D$ and its corresponding U(1) bond dimension $\tilde D$, thus our largest applied $D=2048$ is equivalent to $\tilde{D}\simeq 12000$. 
Respecting SU(2) provides a significant computational and memory benefit in these cases.

\subsubsection{The \ce{Fe1S4}, $n=1$ case}
\begin{figure}
    \centering    \includegraphics[width=0.5\textwidth]{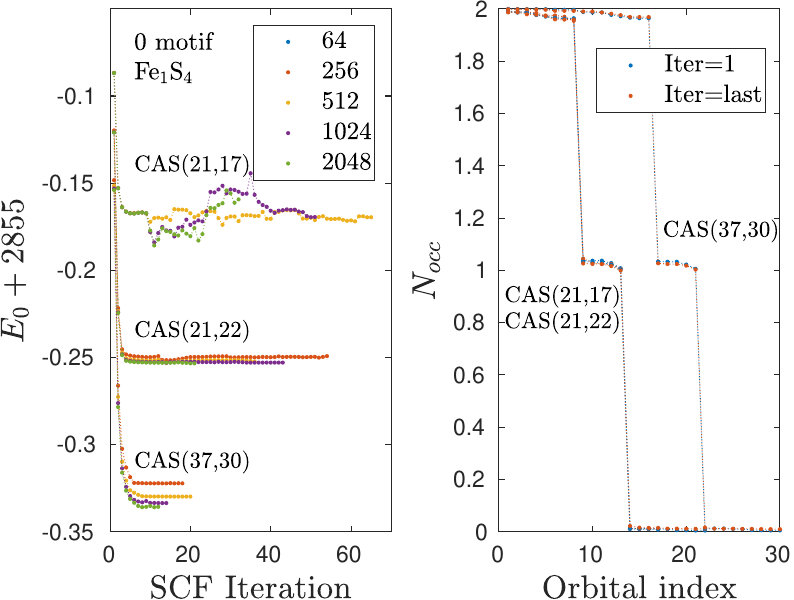}
\caption{(left) Convergence profiles of the DMRG-SCF method for the \ce{Fe_1S_4} cluster for various bond dimension values and for three different active spaces. (right) occupation number distribution of active space orbitals calculated by the first and last DMRG-SCF iterations.}
    \label{fig:0fe2s2}
\end{figure}

In Fig.~\ref{fig:0fe2s2}(left) we show the convergence of the DMRG-SCF procedure for three different active spaces of the \ce{Fe1S4} model constructed from orbitals given in Table.~\ref{tab:fe2s2_cas}, labeled by `A', `B', and `C', i.e. for 
CAS(21,17), CAS(21,22), and CAS(37,30), respectively, to study the effect of proper CAS selection.
It is obvious that for a given CAS, a lower energy is reached with increasing $D$ values as has also been shown for the PAH series. For too small $D=64$, however, we found a highly oscillating profile, indicating that the DMRG wavefunction was not accurate enough (not shown for the larger CAS configurations). 
With regard to the CAS dependence, we confirmed that the energy decreases with increasing CAS size, as expected. 
However, it is important to note that for the smallest CAS, the DMRG-SCF procedure did not converge, even employing a very large bond dimension $D=2048$, characterized by a very small truncation error of the order of $10^{-7}$.
Consequently, the DMRG result can be considered as a full-CI solution, and the numerical instability is due to the insufficiency of the truncated active space, namely due to the exclusion of the 4d orbitals of the \ce{Fe} atom.
We have further confirmed this by additional calculations using different orbital optimization techniques in the ORCA program, i.e., applying SuperCI+DIIS with level shift, instead of the default SuperCI(PT). Here, we got $\sim$6 m$E_\mathrm{h}$ lower energy, but still experienced convergence issues. More importantly, orbitals corresponding to the 4p orbitals of Fe have been replaced by the 4d orbitals of Fe, signaling the poor choice of the selected CAS, and the importance of 4d orbitals.
In contrast to this, for the larger active space ``B'', including the 4d orbitals of the \ce{Fe} atom as well, we found a stable convergence, 
but still, a very large $D$ value had to be enforced to reduce the number of SCF iteration steps (for $D=2048$ the optimization procedure terminated after 23 SCF iterations).
In contrast to this, for the largest active space ``C'', which includes all chemically relevant orbitals of the \ce{Fe} and \ce{S} atoms except the sigma bonds of the \ce{SH} ligands, we found a very fast and stable convergence. The DMRG-SCF procedure terminated after 13-24 iterations. According to tight error settings in the CASSCF module of ORCA, the change in energy in the last three SFC iterations for $D\ge 512$ was in the range of $10^{-4}$.
The different numbers of data points shown for the various $D$ values refer to converged DMRG-SCF results.

In Fig.~\ref{fig:0fe2s2}(right) we show the occupation number distribution for the largest active space CAS(37,30), which 
includes the additional 3s orbitals of the \ce{S} and 3s and 3p orbitals of \ce{Fe} atom.
The five singly occupied d-orbitals of the Fe atom are well visible in the figure, remaining orbitals are either almost fully occupied or empty. 
The obtained profiles for the smaller active spaces are also shown for the sake of completeness, but they almost overlap in the given scale. It is clearly visible, however, that the additional occupied orbitals between the largest two active spaces are characterized with an occupation number larger than 1.99. 
Regarding the effect of orbital optimization on the occupation number distribution, we note that the changes in the profiles measured between the first and last DMRG-SCF iterations have been found to be of order $10^{-2}$ (see the small differences between the red and blue data points for each CAS space). 
Moreover, the plateau of the d orbitals has been preserved; thus, the initial configuration has not been altered by the SCF procedure. This, together with the small changes in the occupation number profile confirms that this system is single reference in nature.

We conclude our analysis, by comparing the best DMRG-SCF energy (see Table \ref{tab:exetime-fe2s2}) to the CCSD(T) energy, E=-2856.16129. This indicates that there is still a discrepancy of 825 mHartree. Therefore, although the SCF procured lead to an improvement of 182 mHartree, additional post-DMRG methods like DMRG-TCC~\cite{Veis-2016,Faulstich-2019a,Liao-2024}, DMRG-RAS-X~\cite{Friesecke-2023,Larsson-2022}, or perturbation theory~\cite{Stolarczyk-1994,Kurashige-2011,Liu-2013,Pulay-2011,Sharma-2014b,Cheng-2022} are required to capture the still missing dynamic correlations. The application of these methods to this and similar systems for recovering these final missing dynamic correlations will be part of a future publication.
Finally, we highlight the fact that DMRG-SCF provides the best variational reference energy, and can be systematically improved with increasing bond dimension and CAS sizes.

\subsubsection{The one \ce{Fe2S2} motif, $n=2$ case}
\begin{figure}
    \centering    \includegraphics[width=0.5\textwidth]{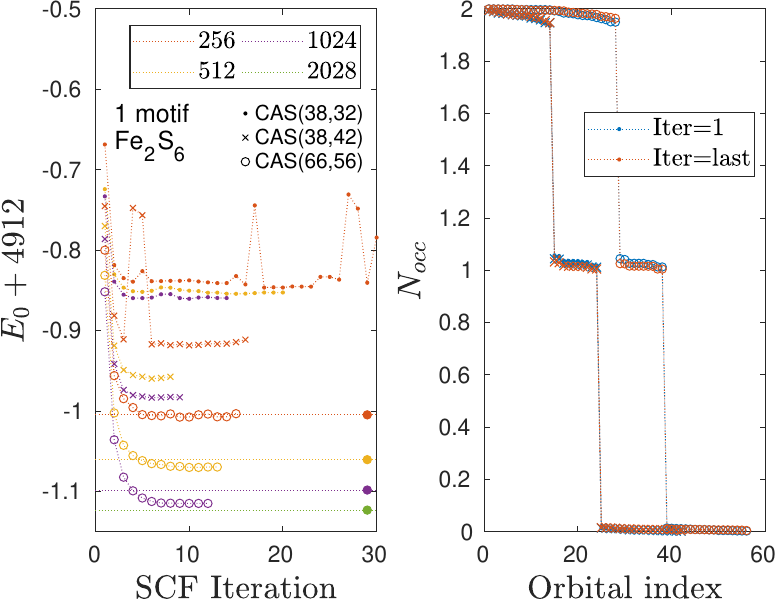}
\caption{Similar to Fig.~\ref{fig:0fe2s2} but for the \ce{Fe2S6H4^{-2}} cluster corresponding to one \ce{Fe2S2} motif, parametrized with $n=1$.
Symbols filled circle
together with horizontal dotted lines stand for post DMRG-
SCF energy values obtained for various $D=\{256,512,1024,2048\}$ values using the
basis optimized with $D_{\rm opt} = 256$. 
}
    \label{fig:1fe2s2}
\end{figure}

A similar analysis for $n=2$, i.e., for \ce{Fe2S6H4^{-2}} corresponding to one \ce{Fe2S2} motif has also been performed, leading to a very similar conclusion. Our results are summarized in Fig.~\ref{fig:1fe2s2}. 
Again, we obtained faster and more stable convergence with increasing CAS sizes and $D$ values. For the smallest CAS the oscillating profile is due to the insufficiency of the active space, while for configuration "B" even $D=256$ was not enough to reach stable convergence.
Therefore, for more complex systems, larger bond dimensions in the SCF procedure are mandatory, which ultimately calls for the hybrid CPU-GPU parallel implementation.
The importance of the inclusion of the 4d orbitals of the Fe atoms in the active spaces is reflected by the more stable convergence profiles for the two larger CASs.
Regarding the quality of the optimized basis obtained via truncated bond dimension, we have performed again post-DMRG-SCF calculations with $D=256,512,1024,2048$ using the basis optimized with $D_{\rm opt}=256$. 
As can be seen, results shown by filled circles together with dotted lines are very close to those where the basis has been optimized via larger, corresponding, bond dimension values.
However, for this more complex molecular system, the basis error gets more pronounced with increasing $D$, that for $D=1024$ we found to be $\delta_E(D,D_{\rm opt}) \simeq 16$ mHartree. This is beyond chemical accuracy and indicates that larger bond dimensions are required in the SCF procedure for more complex problems to reduce basis error.
On the other hand, using post-DMRG-SCF data sets and a proper extrapolation via $1/D$ or the truncation error allows us to
reach chemical accuracy.

Regarding the occupation number profiles, now the ten d-orbitals of the two Fe atoms are singly occupied as shown in Fig.~\ref{fig:1fe2s2}(right).
The difference between the two larger active spaces is marginal except that the Fermi surface is shifted. The occupation numbers of all added orbitals have again been found to be larger than 1.99. 
Since these almost fully occupied orbitals have only marginal contributions to correlation effects, for larger $n$ values we performed calculations employing only CAS configuration "B".

We close our analysis by remarking on the difficulty of the underlying calculations. 
We have found that, due to the close-lying eigenstates, the DMRG-SCF is very sensitive to accuracy settings. Therefore, if a not fully converged DMRG reference wave function is used, the SCF procedure can lead to unreliable results. For example, for a higher residual error in the L\'anczos diagonalization (${10^{-6}}$) or for not properly optimized orbital ordering, or not enough sweeps we could not reproduce the plateau of the d orbitals, and usually 3-4 scattered data points of occupation numbers appeared in the range of 1.4-1.8 and 0.2-0.6 even employing a large $D=2048$.
Moreover, performing post-DMRG-SCF calculations, starting with a badly converged SCF result, despite using very large bond dimension values up to $D~\sim6000$, the correct occupation number profile could not be recovered. Therefore, in practice, it is not enough to monitor the convergence of the energy alone, but further properties of the wave function must also be analyzed. This follows from the fact that in a fixed-rank MPS manifold, several local minima with close energy values can correspond to wave functions with different properties~\cite{Boguslawski-2011}.

For completeness, we also remark that if the d-orbitals were not localized initially, then the proper shape of the plateau could have been obtained with very large $D\ge2048$ bond dimension values only even for the smallest active space. This also highlights the importance of the preparation of initial orbitals and their significant effect on the convergence speed and quality of the final result of DMRG-SCF.

\subsubsection{Several \ce{Fe2S2} motifs with $n\ge3$}

\begin{figure}
    \centering    \includegraphics[width=0.5\textwidth]{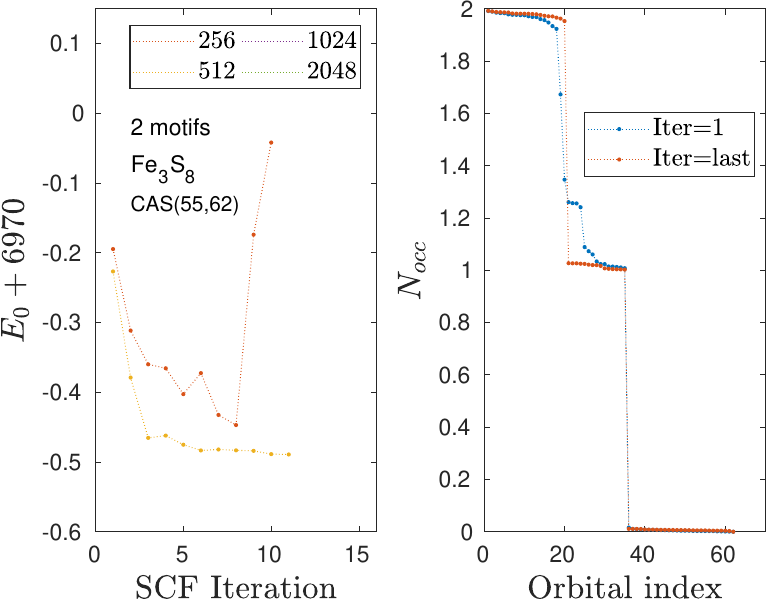}
\caption{Similar to Fig.~\ref{fig:0fe2s2} but for the \ce{Fe3S8H4^{3-}} 
%and \ce{Fe4S10H4^{4-}} 
cluster corresponding to two 
\ce{Fe2S2} motifs parametrized with $n=3$.
}
    \label{fig:4fe2s2}
\end{figure}

For larger $n$ values, i.e., for more \ce{Fe2S2} motifs, we have obtained similar results employing CAS configuration ``B''. Convergence profiles for $n=3$, i.e., for two motifs, are shown in Fig.~\ref{fig:4fe2s2}.
Here, for $n=3$ corresponding to \ce{Fe3S8H4^{3-}} using CAS(55,62) we recovered the 15 singly occupied d-orbitals, after careful orbital ordering optimization. For tutorial purposes, we show the result for $n=3$ starting DMRG-SCF from a not fully optimized ordering configuration with non fully converged wave function. As can be seen, even if the initial wave function was inadequate (see occupation number distribution at the plateau), DMRG-SCF can correct it, but as discussed before, there is no guarantee for that. Using a fully optimized initial DMRG configuration, we have a higher chance of getting the expected distribution of occupation numbers, which are usually preserved via the DMRG-SCF procedure. The main difficulty is due to the fact that only the total spin is conserved, but not the spin states of the individual iron atoms. 
Moreover, due to the larger computational complexity of the problem, larger bond dimensions are mandatory. As can be seen, for $D=256$ DMRG-SCF can even loose the target after a few iteration steps when the large CAS error starts to influence the stability of the SCF procedure. Such non converging profile can be seen in  Fig.~\ref{fig:4fe2s2}.

For $n=4$, for \ce{Fe4S10H4^{4-}}, we faced further serious difficulties even to recover the plateau of the 20 singly occupied d-orbitals on the applied CAS(72,82) model space. In spite of performing orbital optimizations via a large number of DMRG sweeps up to 50-60 together with bond dimensions up to 1024, we have always found a few scattered data points in the range of 1.4-1.8 and 0.2-0.6 of occupation. Therefore, finding the correct target state for DMRG is a hard task as the number of singly occupied orbitals increases, for which open shell problems, treated with large active spaces, further investigation is necessary. A promising direction would be
a combination with 
restricted open-shell configuration interaction singles (ROCIS) family of methods~\cite{Leyser-2024}. 

To provide further insights in Table~\ref{tab:exetime-fe2s2}, we summarize the computational time via fixed $D$ values for various parameter sets.
\begin{table}[t]
  \centering
\begin{tabular}{l|c|c|c|c|c|c|c|r}
\hline
 \hline
$\tilde N_{\rm e}$ & $\tilde N$ & CAS & $D_{\rm min}$ & $D_{\rm max}$ & ${\tilde D}_{\rm max}$ & $t_{\rm A100}$ & $t_{\rm H100}$ & Energy\\
 \hline
 95 & 135 & (37,30) & 256  & 256  & 1551  & 12.7 h  & 7.3 h& -2855.3221\\ 
 95 & 135 & (37,30) & 512 & 512 & 3164  & 26.2 h & 13.5 h& -2855.3297\\ 
  95 & 135 & (37,30) & 1024 & 1024 & 6218  & 29.5 h & 17.4 h& -2855.3334\\ 
    95 & 135 & (37,30) & 2048 & 2048 & 12478  & 54.1 h & 33.6 h& -2855.3356\\
 \hline
 154 & 214 & (66,56) & 256 & 256 & 769 & 23.9 h & 14.3 h& -4913.0073 \\ 
  154 & 214 & (66,56) & 512 & 512 & 1591 & 50.7 h & 23.1 h & -4913.0699 \\ 
  154 & 214 & (66,56) & 1024 & 1024 & 3197 & 80.6 h & 47.4 h & -4913.1147\\ 
 \hline
 213 & 293 & (55,62) & 256 & 256 & 1671 & NaN & NaN & NaN\\
 213 & 293 & (55,62) & 512 & 512 & 3467 & 123 h & 55.7 h& -6970.4836\\
 \hline
 \hline
\end{tabular}
\caption{Total number of electrons, $\tilde N_{\rm e}$, and orbitals, $\tilde N$, of the full orbital space, size of selected CAS, 
minimum DMRG bond dimension, $D_{\rm min}$, 
maximum DMRG bond dimension, $D_{\rm max}$, 
corresponding maximum DMRG U(1) bond dimension, ${\tilde D}_{\rm max}$, 
and total time for the eight A100 and H100 GPU accelerated DMRG-SCF for the 
Iron-sulfur clusters with increasing number of \ce{Fe2S2} motifs, $n$.
}
\label{tab:exetime-fe2s2}
\end{table}
 
\section{Conclusion}

In this work, we have presented an efficient orbital optimization procedure utilizing the massively parallel spin adapted ab initio density matrix renormalization group (DMRG) method via the complete active space self consistent field (CASSCF) framework on high performance computing infrastructures building on state-of-the-art hardware and software technologies.
Our DMRG-SCF method employs the message passing interface (MPI) based parallel implementation of the SCF procedure accessible via the Orca program package~\cite{orca-2020} which works in a perfect synergy with our hybrid CPU-multiGPU DMRG implementation~\cite{Menczer-2023a,Menczer-2023b,Menczer-2024a,Menczer-2024c}. 
Substantial speed ups compared to existing CPU-based implementations is also due to the utilization of a GPU accelerated method to compute
the one- and two-particle reduced density matrices, which we have put to use for running CASSCF calculations at large bond dimensions to obtain stable convergence with high accuracy and at unprecedented AS sizes of up to 82 electrons in 82 orbitals. 

Our results for the polycyclic
aromatic hydrocarbons (PAHs) obtained with various DMRG bond dimension values demonstrate the crucial need for accurate CAS solutions to reduce the number of SCF iterations and to reach converged results with an accuracy in the order of milliHartree. Moreover, we provided a detailed error analysis as a function of the DMRG bond dimension and DMRG truncation error. The quality of the optimized basis obtained with lower accuracy, i.e., larger truncation error, has been analyzed, indicating that basis error with larger active space sizes increases systematically. Nevertheless, by performing extrapolations with inverse bond dimension as part of a post-SCF DMRG procedure, one can approximate the truncation-free solution. On the other hand, much larger bond dimensions are needed if a non-optimized basis is used.

Our analysis on iron-sulfur clusters reveals the importance of proper CAS selection and constraints on minimal CAS sizes to achieve stable convergence via the DMRG-SCF procedure. Numerical challenges related to such open-shell systems have been discussed in detail, along with the reported higher computational complexity and total wall time.
Nevertheless, the singly occupied d-orbitals of the Fe atoms have been recovered, and the corresponding plateau in the occupation number profile has been argued to be an important quantity to monitor besides ground state energy. 
A natural extension of our work would be towards Fe(II) porphyrin~\cite{Antalik-2020}, which is a particularly interesting system where 
treating all $\pi$ orbitals as well as selected sigma electrons, double d-shell, and semi-core orbitals may finally lead to a converged result.

Regarding computational time, we have also presented the measured wall time of full DMRG-SCF protocol on NVIDIA DGX-A100 and DGX-H100 hardware. These indicate that by considering large active spaces for systems composed of several hundreds of electrons over thousands of orbitals, the CASSCF-based orbital optimization can be revolutionized by taking advantage of the underlying computational power in AI accelerators available via graphics process units. In addition, for intermediate bond dimension values, i.e., for $D\le2000$, we measured a factor of 1.6-2.3 speedup by switching from an A100 to an H100 node. For larger $D$ values, this speedup can even be larger based on our previous DMRG CAS benchmark calculations~\cite{Menczer-2024b}.

Finally, we remark that results presented in the current work have been obtained without employing the dynamically extended active space (DEAS) procedure~\cite{Legeza-2003b,Barcza-2011} since its SU(2) spin-adapted version is not available yet. The lack of such an initialization protocol is the reason for the high number of DMRG sweeps required in the DMRG-SCF approach. This is, however, part of our current developments, thus a factor of two to three reduction in sweeping is expected just like in our U(1) implementation. Further reduction in wall time is also expected by synchronizing the various error terms in DMRG and SCF even dynamically. Therefore, in the near future, our hybrid CPU-multiGPU DMRG code interfaced with the ORCA program package has the potential to target complex strongly correlated molecular clusters, including several transition metal centers, as a routinely applied daily method.

\section*{Acknowledgments}
This work has been supported by the Hungarian National Research, Development and Innovation Office (NKFIH) through Grant Nos.~K134983, PD146265 and TKP2021-NVA-04, by the Quantum Information National Laboratory
of Hungary. \"O.L. acknowledges financial support
by the Hans Fischer Senior Fellowship programme funded by the Technical University
of Munich – Institute for Advanced Study. \"O.L. and S.S.X. acknowledge support from 
the Center for Scalable and Predictive methods
for Excitation and Correlated phenomena (SPEC),
funded as part of the Computational Chemical Sciences Program FWP 70942 by the U.S. Department of Energy
(DOE), Office of Science, Office of Basic Energy Sciences, Division of Chemical Sciences, Geosciences, and Biosciences at Pacific Northwest National Laboratory.
M.A.W. has also been supported by the Janos
Bolyai Research Scholarship of the Hungarian Academy of Sciences.
The simulations were performed on the national supercomputer HPE Apollo Hawk at the High Performance Computing Center Stuttgart (HLRS) under the grant number MPTNS/44246, at the Wigner Scientific Computing Laboratory (WSCLAB), and on Google Cloud.

%----------------------------------------------------------------------------------------
%	 REFERENCES
%----------------------------------------------------------------------------------------

%\normalem
%\bibliographystyle{apsrev4-1}
%\bibliographystyle{unsrt}
%
%\bibliography{paper_new}

\begin{thebibliography}{100}

    \bibitem{orca-2020}
    Frank Neese, Frank Wennmohs, Ute Becker, and Christoph Riplinger.
    \newblock The {ORCA} quantum chemistry program package.
    \newblock {\em The Journal of Chemical Physics}, 152(22):224108, June 2020.
    \newblock \_eprint:
      https://pubs.aip.org/aip/jcp/article-pdf/doi/10.1063/5.0004608/16740678/224108\_1\_online.pdf.
    
    \bibitem{nwchem-2020}
    E.~et~al. Apr\`a.
    \newblock {NWChem}: Past, present, and future.
    \newblock {\em The Journal of Chemical Physics}, 152(18):184102, 2020.
    
    \bibitem{molpro-2012}
    Hans-Joachim Werner, Peter~J. Knowles, Gerald Knizia, Frederick~R. Manby, and
      Martin Schütz.
    \newblock Molpro: a general-purpose quantum chemistry program package.
    \newblock {\em WIREs Computational Molecular Science}, 2(2):242--253, 2012.
    
    \bibitem{Gaussian-2016}
    M.~J.~Frisch et~al.
    \newblock Gaussian˜16 {R}evision {C}.01, 2016.
    \newblock Gaussian Inc. Wallingford CT.
    
    \bibitem{qchem-2015}
    Yihan~Shao et~al.
    \newblock Advances in molecular quantum chemistry contained in the {Q-Chem} 4
      program package.
    \newblock {\em Molecular Physics}, 113(2):184--215, 2015.
    
    \bibitem{ADF-2001}
    G.~te~Velde, F.~M. Bickelhaupt, E.~J. Baerends, C.~Fonseca~Guerra, S.~J.~A. van
      Gisbergen, J.~G. Snijders, and T.~Ziegler.
    \newblock Chemistry with adf.
    \newblock {\em J. Comput. Chem.}, 22(9):931--967, 2001.
    
    \bibitem{VASP-1996}
    G.~Kresse and J.~Furthm\"uller.
    \newblock Efficient iterative schemes for \textit{ab initio} total-energy
      calculations using a plane-wave basis set.
    \newblock {\em Phys. Rev. B}, 54:11169--11186, Oct 1996.
    
    \bibitem{pyscf-2020}
    Qiming et~al. Sun.
    \newblock Recent developments in the {PySCF} program package.
    \newblock {\em The Journal of Chemical Physics}, 153(2):024109, July 2020.
    \newblock \_eprint:
      https://pubs.aip.org/aip/jcp/article-pdf/doi/10.1063/5.0006074/16722275/024109\_1\_online.pdf.
    
    \bibitem{terachem-2021}
    Stefan Seritan, Christoph Bannwarth, Bryan~S. Fales, Edward~G. Hohenstein,
      Christine~M. Isborn, Sara I.~L. Kokkila-Schumacher, Xin Li, Fang Liu, Nathan
      Luehr, James~W. Snyder~Jr., Chenchen Song, Alexey~V. Titov, Ivan~S. Ufimtsev,
      Lee-Ping Wang, and Todd~J. Martínez.
    \newblock {TeraChem}: A graphical processing unit-accelerated electronic
      structure package for large-scale ab initio molecular dynamics.
    \newblock {\em WIREs Computational Molecular Science}, 11(2):e1494, 2021.
    
    \bibitem{quantum_espresso-2009}
    Paolo Giannozzi, Stefano Baroni, Nicola Bonini, Matteo Calandra, Roberto Car,
      Carlo Cavazzoni, Davide Ceresoli, Guido~L Chiarotti, Matteo Cococcioni,
      Ismaila Dabo, Andrea~Dal Corso, Stefano de~Gironcoli, Stefano Fabris, Guido
      Fratesi, Ralph Gebauer, Uwe Gerstmann, Christos Gougoussis, Anton Kokalj,
      Michele Lazzeri, Layla Martin-Samos, Nicola Marzari, Francesco Mauri,
      Riccardo Mazzarello, Stefano Paolini, Alfredo Pasquarello, Lorenzo Paulatto,
      Carlo Sbraccia, Sandro Scandolo, Gabriele Sclauzero, Ari~P Seitsonen,
      Alexander Smogunov, Paolo Umari, and Renata~M Wentzcovitch.
    \newblock Quantum espresso: a modular and open-source software project for
      quantum simulations of materials.
    \newblock {\em Journal of Physics: Condensed Matter}, 21(39):395502, sep 2009.
    
    \bibitem{quantum_espresso-2017}
    P~Giannozzi, O~Andreussi, T~Brumme, O~Bunau, M~Buongiorno Nardelli, M~Calandra,
      R~Car, C~Cavazzoni, D~Ceresoli, M~Cococcioni, N~Colonna, I~Carnimeo, A~Dal
      Corso, S~de~Gironcoli, P~Delugas, R~A DiStasio, A~Ferretti, A~Floris,
      G~Fratesi, G~Fugallo, R~Gebauer, U~Gerstmann, F~Giustino, T~Gorni, J~Jia,
      M~Kawamura, H-Y Ko, A~Kokalj, E~Küçükbenli, M~Lazzeri, M~Marsili,
      N~Marzari, F~Mauri, N~L Nguyen, H-V Nguyen, A~Otero de-la Roza, L~Paulatto,
      S~Poncé, D~Rocca, R~Sabatini, B~Santra, M~Schlipf, A~P Seitsonen,
      A~Smogunov, I~Timrov, T~Thonhauser, P~Umari, N~Vast, X~Wu, and S~Baroni.
    \newblock Advanced capabilities for materials modelling with quantum espresso.
    \newblock {\em Journal of Physics: Condensed Matter}, 29(46):465901, oct 2017.
    
    \bibitem{quantum_espresso-2020}
    Paolo Giannozzi, Oscar Baseggio, Pietro Bonfà, Davide Brunato, Roberto Car,
      Ivan Carnimeo, Carlo Cavazzoni, Stefano de~Gironcoli, Pietro Delugas,
      Fabrizio Ferrari~Ruffino, Andrea Ferretti, Nicola Marzari, Iurii Timrov,
      Andrea Urru, and Stefano Baroni.
    \newblock Quantum espresso toward the exascale.
    \newblock {\em The Journal of Chemical Physics}, 152(15):154105, 04 2020.
    
    \bibitem{Pederson-2023}
    Ryan Pederson, John Kozlowski, Ruyi Song, Jackson Beall, Martin Ganahl, Markus
      Hauru, Adam G.~M. Lewis, Yi~Yao, Shrestha~Basu Mallick, Volker Blum, and
      Guifre Vidal.
    \newblock Large scale quantum chemistry with tensor processing units.
    \newblock {\em Journal of Chemical Theory and Computation}, 19(1):25--32, 2023.
    \newblock PMID: 36508260.
    
    \bibitem{Li-Manni-2018}
    Giovanni Li~Manni and Ali Alavi.
    \newblock Understanding the mechanism stabilizing intermediate spin states in
      fe(ii)-porphyrin.
    \newblock {\em The Journal of Physical Chemistry A}, 122(22):4935--4947, 2018.
    \newblock PMID: 29595978.
    
    \bibitem{Reiher-2009}
    Markus Reiher.
    \newblock A theoretical challenge: Transition-metal compounds.
    \newblock {\em CHIMIA}, 63:140, 2009.
    
    \bibitem{Li-2019}
    Zhendong Li, Junhao Li, Nikesh~S. Dattani, C.~J. Umrigar, and Garnet Kin-Lic
      Chan.
    \newblock The electronic complexity of the ground-state of the {FeMo} cofactor
      of nitrogenase as relevant to quantum simulations.
    \newblock {\em The Journal of Chemical Physics}, 150(2):024302, 01 2019.
    
    \bibitem{Evangelista-2018}
    Francesco~A. Evangelista.
    \newblock Perspective: Multireference coupled cluster theories of dynamical
      electron correlation.
    \newblock {\em The Journal of Chemical Physics}, 149(3):030901, 2018.
    
    \bibitem{Yudong-2019}
    Yudong Cao, Jonathan Romero, Jonathan~P. Olson, Matthias Degroote, Peter~D.
      Johnson, M{\'a}ria Kieferov{\'a}, Ian~D. Kivlichan, Tim Menke, Borja
      Peropadre, Nicolas P.~D. Sawaya, Sukin Sim, Libor Veis, and Al{\'a}n
      Aspuru-Guzik.
    \newblock Quantum chemistry in the age of quantum computing.
    \newblock {\em Chemical Reviews}, 119(19):10856--10915, 2019.
    \newblock PMID: 31469277.
    
    \bibitem{Bauer-2020}
    Bela Bauer, Sergey Bravyi, Mario Motta, and Garnet Kin-Lic Chan.
    \newblock Quantum algorithms for quantum chemistry and quantum materials
      science.
    \newblock {\em Chemical Reviews}, 120(22):12685--12717, 2020.
    \newblock PMID: 33090772.
    
    \bibitem{Baiardi-2020}
    Alberto Baiardi and Markus Reiher.
    \newblock The density matrix renormalization group in chemistry and molecular
      physics: Recent developments and new challenges.
    \newblock {\em The Journal of Chemical Physics}, 152(4):040903, 2020.
    
    \bibitem{Affleck-1987}
    Ian Affleck, Tom Kennedy, Elliott~H. Lieb, and Hal Tasaki.
    \newblock Rigorous results on valence-bond ground states in antiferromagnets.
    \newblock {\em Phys. Rev. Lett.}, 59:799--802, Aug 1987.
    
    \bibitem{White-1992b}
    Steven~R. White.
    \newblock Density matrix formulation for quantum renormalization groups.
    \newblock {\em Phys. Rev. Lett.}, 69:2863--2866, November 1992.
    
    \bibitem{White-1999}
    Steven~R. White and Richard~L. Martin.
    \newblock Ab initio quantum chemistry using the density matrix renormalization
      group.
    \newblock {\em The Journal of Chemical Physics}, 110(9):4127--4130, 1999.
    
    \bibitem{Chan-2002a}
    Garnet Kin-Lic Chan and Martin Head-Gordon.
    \newblock Highly correlated calculations with a polynomial cost algorithm: A
      study of the density matrix renormalization group.
    \newblock {\em The Journal of Chemical Physics}, 116(11):4462--4476, 2002.
    
    \bibitem{Legeza-2003a}
    {\"O}.~Legeza, J.~R\"oder, and B.~A. Hess.
    \newblock Controlling the accuracy of the density-matrix renormalization-group
      method: The dynamical block state selection approach.
    \newblock {\em Phys. Rev. B}, 67:125114, Mar 2003.
    
    \bibitem{Legeza-2008}
    {\"O}.~Legeza, R.M. Noack, J.~S\'olyom, and L.~Tincani.
    \newblock Applications of quantum information in the density-matrix
      renormalization group.
    \newblock In H.~Fehske, R.~Schneider, and A.~Weisse, editors, {\em
      Computational Many-Particle Physics}, volume 739 of {\em Lecture Notes in
      Physics}, pages 653--664. Springer, Berlin, Heidelberg, 2008.
    
    \bibitem{Chan-2008}
    Garnet Kin-Lic Chan, Jonathan~J. Dorando, Debashree Ghosh, Johannes Hachmann,
      Eric Neuscamman, Haitao Wang, and Takeshi Yanai.
    \newblock An introduction to the density matrix renormalization group ansatz in
      quantum chemistry.
    \newblock In Stephen Wilson, Peter~J. Grout, Jean Maruani, Gerardo
      Delgado-Barrio, and Piotr Piecuch, editors, {\em Frontiers in Quantum Systems
      in Chemistry and Physics}, volume~18 of {\em Progress in Theoretical
      Chemistry and Physics}. Springer, Netherlands, 2008.
    
    \bibitem{Yanai-2009}
    Takeshi Yanai, Yuki Kurashige, Debashree Ghosh, and Garnet Kin-Lic Chan.
    \newblock Accelerating convergence in iterative solution for large-scale
      complete active space self-consistent-field calculations.
    \newblock {\em International Journal of Quantum Chemistry}, 109(10):2178--2190,
      2009.
    
    \bibitem{Kurashige-2009}
    Yuki Kurashige and Takeshi Yanai.
    \newblock High-performance ab initio density matrix renormalization group
      method: Applicability to large-scale multireference problems for metal
      compounds.
    \newblock {\em The Journal of Chemical Physics}, 130(23):234114, 2009.
    
    \bibitem{Marti-2010c}
    Konrad~H. Marti and Markus Reiher.
    \newblock The density matrix renormalization group algorithm in quantum
      chemistry.
    \newblock {\em Zeitschrift f\"ur Physikalische Chemie}, 224:583--599, 2010.
    
    \bibitem{Wouters-2014a}
    Sebastian Wouters, Ward Poelmans, Paul~W. Ayers, and Dimitri~Van Neck.
    \newblock {CheMPS2}: A free open-source spin-adapted implementation of the
      density matrix renormalization group for ab initio quantum chemistry.
    \newblock {\em Computer Physics Communications}, 185(6):1501 -- 1514, 2014.
    
    \bibitem{Szalay-2015a}
    {\relax Sz}il{\'a}rd {\relax Sz}alay, Max Pfeffer, Valentin Murg, Gergely
      Barcza, Frank Verstraete, Reinhold Schneider, and {\"O}rs Legeza.
    \newblock Tensor product methods and entanglement optimization for ab initio
      quantum chemistry.
    \newblock {\em Int. J. Quantum Chem.}, 115(19):1342--1391, 2015.
    
    \bibitem{Chan-2016}
    Garnet Kin-Lic Chan, Anna Keselman, Naoki Nakatani, Zhendong Li, and Steven~R.
      White.
    \newblock Matrix product operators, matrix product states, and ab initio
      density matrix renormalization group algorithms.
    \newblock {\em The Journal of Chemical Physics}, 145(1):014102, 2016.
    
    \bibitem{Cheng-2022}
    Yifan Cheng, Zhaoxuan Xie, and Haibo Ma.
    \newblock Post-density matrix renormalization group methods for describing
      dynamic electron correlation with large active spaces.
    \newblock {\em The Journal of Physical Chemistry Letters}, 13(3):904--915,
      2022.
    
    \bibitem{Schollwock-2005}
    Ulrich Schollw\"ock.
    \newblock The density-matrix renormalization group.
    \newblock {\em Rev. Mod. Phys.}, 77:259--315, Apr 2005.
    
    \bibitem{Schollwock-2011}
    Ulrich Schollw\"ock.
    \newblock The density-matrix renormalization group in the age of matrix product
      states.
    \newblock {\em Annals of Physics}, 326(1):96 -- 192, 2011.
    \newblock January 2011 Special Issue.
    
    \bibitem{Verstraete-2004}
    F.~Verstraete and J.~I. Cirac.
    \newblock Renormalization algorithms for quantum-many body systems in two and
      higher dimensions, 2004.
    
    \bibitem{Corboz-2009}
    Philippe Corboz and Guifr\'e Vidal.
    \newblock Fermionic multiscale entanglement renormalization ansatz.
    \newblock {\em Phys. Rev. B}, 80:165129, Oct 2009.
    
    \bibitem{Vidal-2007}
    G.~Vidal.
    \newblock Entanglement renormalization.
    \newblock {\em Phys. Rev. Lett.}, 99:220405, Nov 2007.
    
    \bibitem{Verstraete-2023}
    Frank Verstraete, Tomotoshi Nishino, Ulrich Schollw{\"o}ck, Mari~Carmen
      Ba{\~n}uls, Garnet~K Chan, and Miles~E Stoudenmire.
    \newblock Density matrix renormalization group, 30 years on.
    \newblock {\em Nature Reviews Physics}, pages 1--4, 2023.
    
    \bibitem{Siegbahn-1980}
    Per Siegbahn, Anders Heiberg, Bj{\"o}rn Roos, and Bernard Levy.
    \newblock A comparison of the super-{CI} and the {N}ewton-{R}aphson scheme in
      the complete active space {SCF} method.
    \newblock {\em Physica Scripta}, 21(3-4):323, 1980.
    
    \bibitem{Roos-1980}
    Bj{\"o}rn~O Roos, Peter~R Taylor, and Per~EM Sigbahn.
    \newblock A complete active space {SCF} method ({CASSCF}) using a density
      matrix formulated super-{CI} approach.
    \newblock {\em Chemical Physics}, 48(2):157--173, 1980.
    
    \bibitem{Siegbahn-1981}
    Per~EM Siegbahn, Jan Alml{\"o}f, Anders Heiberg, and Bj{\"o}rn~O Roos.
    \newblock The complete active space {SCF} ({CASSCF}) method in a
      {N}ewton--{R}aphson formulation with application to the {HNO} molecule.
    \newblock {\em The Journal of Chemical Physics}, 74(4):2384--2396, 1981.
    
    \bibitem{Menczer-2023b}
    Andor Menczer and \"{O}rs Legeza.
    \newblock Tensor network state algorithms on {AI} accelerators.
    \newblock {\em Journal of Chemical Theory and Computation}, 20(20):8897--8910,
      2024.
    
    \bibitem{Xiang-2024}
    Chunyang Xiang, Weile Jia, Wei-Hai Fang, and Zhendong Li.
    \newblock A distributed multi-{GPU} ab initio density matrix renormalization
      group algorithm with applications to the {P}-cluster of nitrogenase.
    \newblock {\em Journal of Chemical Theory and Computation}, 20(2):775–786,
      2024.
    
    \bibitem{Zgid-2008c}
    Dominika Zgid and Marcel Nooijen.
    \newblock The density matrix renormalization group self-consistent field
      method: Orbital optimization with the density matrix renormalization group
      method in the active space.
    \newblock {\em The Journal of Chemical Physics}, 128(14):144116, 2008.
    
    \bibitem{Liu-2013}
    Fengyi Liu, Yuki Kurashige, Takeshi Yanai, and Keiji Morokuma.
    \newblock Multireference ab initio density matrix renormalization group
      {(DMRG)-CASSCF} and {DMRG-CASPT2} study on the photochromic ring opening of
      spiropyran.
    \newblock {\em Journal of Chemical Theory and Computation}, 9(10):4462--4469,
      2013.
    
    \bibitem{Wouters-2014b}
    Sebastian Wouters, Thomas Bogaerts, Pascal Van Der~Voort, Veronique
      Van~Speybroeck, and Dimitri Van~Neck.
    \newblock Communication: {DMRG-SCF} study of the singlet, triplet, and quintet
      states of {oxo-Mn(Salen)}.
    \newblock {\em The Journal of Chemical Physics}, 140(24):241103, 2014.
    
    \bibitem{Ding-2023}
    Lexin Ding, Stefan Knecht, and Christian Schilling.
    \newblock Quantum information-assisted complete active space optimization
      ({QICAS}).
    \newblock {\em The Journal of Physical Chemistry Letters},
      14(49):11022–11029, December 2023.
    
    \bibitem{Menczer-2023a}
    Andor Menczer and {\"O}rs Legeza.
    \newblock Massively parallel tensor network state algorithms on hybrid
      {CPU}-{GPU} based architectures.
    \newblock {\em Journal of Chemical Theory and Computation}, 21(4):1572--1587,
      2025.
    
    \bibitem{Menczer-2024c}
    Andor Menczer and {\"O}rs Legeza.
    \newblock Cost optimized ab initio tensor network state methods: industrial
      perspectives.
    \newblock {\em arXiv:2412.04676}, 2024.
    
    \bibitem{Veis-2016}
    Libor Veis, Andrej Antalik, Jiri Brabec, Frank Neese, {\"O}rs Legeza, and Jiri
      Pittner.
    \newblock Coupled cluster method with single and double excitations tailored by
      matrix product state wave functions.
    \newblock {\em The Journal of Physical Chemistry Letters}, 7(20):4072--4078,
      2016.
    
    \bibitem{Demel-2015}
    O.~Demel, J.~Pittner, and F.~Neese.
    \newblock A local pair natural orbital-based multireference {M}ukherjee's
      coupled cluster method.
    \newblock {\em J. Chem. Theory Comput.}, 11:3104--3114, 2015.
    
    \bibitem{Brabec-2018}
    J.~Brabec, J.~Lang, M.~Saitow, J.~Pittner, F.~Neese, and O.~Demel.
    \newblock Domain-based local pair natural orbital version of {M}ukherjee's
      state-specific coupled cluster method.
    \newblock {\em J. Chem. Theory Comput.}, 14:1370--1382, 2018.
    
    \bibitem{Lang-2019}
    J.~Lang, J.~Brabec, M.~Saitow, J.~Pittner, F.~Neese, and O.~Demel.
    \newblock Perturbative triples correction to domain-based local pair natural
      orbital version of {M}ukherjee's state specific coupled cluster method.
    \newblock {\em Phys. Chem. Chem. Phys.}, 21:5022--5038, 2019.
    
    \bibitem{Lang-2020}
    J.~Lang, A.~Antal{\'{\i}}k, L.~Veis, J.~Brabec, O.~Legeza, and J.~Pittner.
    \newblock Near-linear scaling in {DMRG}-based tailored coupled clusters: An
      implementation of {DLPNO-TCCSD} and {DLPNO-TCCSD(T)}.
    \newblock {\em J. Chem. Theory Comput.}, 16:3028--3040, 2020.
    
    \bibitem{Antalik-2020}
    A.~Antal{\'{\i}}k, D.~Nachtigallov{\'{a}}, R.~Lo, M.~Matou{\v{s}}ek, J.~Lang,
      {\"O}.~Legeza, J.~Pittner, P.~Hobza, and L.~Veis.
    \newblock Ground state of the {Fe(II)}-porphyrin model system corresponds to
      the quintet state: {DFT}, {DMRG-TCCSD} and {DMRG-TCCSD(T)} computations.
    \newblock {\em Phys. Chem. Chem. Phys.}, 22:17033--17037, 2020.
    
    \bibitem{Menczer-2024d}
    Andor Menczer and {\"O}rs Legeza.
    \newblock Petaflops density matrix renormalization group method, unpublished,
      2023.
    
    \bibitem{Kollmar-2019}
    Christian Kollmar, Kantharuban Sivalingam, Benjamin Helmich-Paris, Celestino
      Angeli, and Frank Neese.
    \newblock A perturbation-based super-{CI} approach for the orbital optimization
      of a {CASSCF} wave function.
    \newblock {\em Journal of Computational Chemistry}, 40(14):1463--1470, 2019.
    
    \bibitem{Zgid-2008b}
    Dominika Zgid and Marcel Nooijen.
    \newblock Obtaining the two-body density matrix in the density matrix
      renormalization group method.
    \newblock {\em The Journal of Chemical Physics}, 128(14):144115, 2008.
    
    \bibitem{Khedkar-2019}
    Abhishek Khedkar and Michael Roemelt.
    \newblock Active space selection based on natural orbital occupation numbers
      from n-electron valence perturbation theory.
    \newblock {\em Journal of Chemical Theory and Computation}, 15(6):3522--3536,
      2019.
    \newblock PMID: 31059643.
    
    \bibitem{Stein-2016}
    Christopher~J Stein and Markus Reiher.
    \newblock Automated selection of active orbital spaces.
    \newblock {\em J. Chem. Theory Comput.}, 12(4):1760--1771, 2016.
    
    \bibitem{Elvira-2017}
    Elvira~R. Sayfutyarova, Qiming Sun, Garnet Kin-Lic Chan, and Gerald Knizia.
    \newblock Automated construction of molecular active spaces from atomic valence
      orbitals.
    \newblock {\em Journal of Chemical Theory and Computation}, 13(9):4063--4078,
      2017.
    \newblock PMID: 28731706.
    
    \bibitem{Ostlund-1995}
    Stellan \"Ostlund and Stefan Rommer.
    \newblock Thermodynamic limit of density matrix renormalization.
    \newblock {\em Phys. Rev. Lett.}, 75:3537--3540, Nov 1995.
    
    \bibitem{Mcculloch-2002}
    Ian~P McCulloch and Mikl{\'o}s Gul{\'a}csi.
    \newblock The non-{A}belian density matrix renormalization group algorithm.
    \newblock {\em Europhysics Letters}, 57(6):852, 2002.
    
    \bibitem{Toth-2008}
    A.~I. T\'oth, C.~P. Moca, \"O. Legeza, and G.~Zar\'and.
    \newblock Density matrix numerical renormalization group for non-{A}belian
      symmetries.
    \newblock {\em Phys. Rev. B}, 78:245109, Dec 2008.
    
    \bibitem{Sharma-2012a}
    Sandeep Sharma and Garnet Kin-Lic Chan.
    \newblock Spin-adapted density matrix renormalization group algorithms for
      quantum chemistry.
    \newblock {\em The Journal of Chemical Physics}, 136(12):124121, 2012.
    
    \bibitem{Keller-2016}
    Sebastian Keller and Markus Reiher.
    \newblock Spin-adapted matrix product states and operators.
    \newblock {\em The Journal of Chemical Physics}, 144(13):134101, apr 2016.
    
    \bibitem{Gunst-2019}
    Klaas Gunst, Frank Verstraete, and Dimitri~Van Neck.
    \newblock Three-legged tree tensor networks with {SU(2)} and molecular point
      group symmetry.
    \newblock {\em Journal of Chemical Theory and Computation}, 15(5):2996--3007,
      apr 2019.
    
    \bibitem{Werner-2020su3}
    Mikl{\'o}s~Antal Werner, C{\u{a}}t{\u{a}}lin~Pa{\c{s}}cu Moca, {\"O}rs Legeza,
      and Gergely Zar{\'a}nd.
    \newblock Quantum quench and charge oscillations in the {SU(3)} {H}ubbard
      model: A test of time evolving block decimation with general non-abelian
      symmetries.
    \newblock {\em Physical Review B}, 102(15):155108, 2020.
    
    \bibitem{Menczer-2024a}
    Andor Menczer, Korn\'el Kap\'as, Mikl\'os~Antal Werner, and {\"O}rs Legeza.
    \newblock Two-dimensional quantum lattice models via mode optimized hybrid
      {CPU}-{GPU} density matrix renormalization group method.
    \newblock {\em Phys. Rev. B}, 109:195148, May 2024.
    
    \bibitem{Legeza-1996}
    {\"O}rs Legeza and G\'abor F\'ath.
    \newblock Accuracy of the density-matrix renormalization-group method.
    \newblock {\em Phys. Rev. B}, 53:14349--14358, Jun 1996.
    
    \bibitem{Noack-2005}
    Reinhard~M. Noack and Salvatore~R. Manmana.
    \newblock Diagonalization‐ and numerical renormalization‐group‐based
      methods for interacting quantum systems.
    \newblock {\em AIP Conference Proceedings}, 789(1):93--163, 2005.
    
    \bibitem{Szalay-2015b}
    {\relax Sz}il{\'a}rd {\relax Sz}alay, Max Pfeffer, Valentin Murg, Gergely
      Barcza, Frank Verstraete, Reinhold Schneider, and {\"O}rs Legeza.
    \newblock Tensor product methods and entanglement optimization for ab initio
      quantum chemistry.
    \newblock {\em Int. J. Quantum Chem.}, 115(19):1342--1391, 2015.
    
    \bibitem{Krumnow-2016}
    C.~Krumnow, L.~Veis, {\"O}.~Legeza, and J.~Eisert.
    \newblock Fermionic orbital optimization in tensor network states.
    \newblock {\em Phys. Rev. Lett.}, 117:210402, Nov 2016.
    
    \bibitem{Krumnow-2021}
    C.~Krumnow, L.~Veis, J.~Eisert, and {\"O}.~Legeza.
    \newblock Effective dimension reduction with mode transformations: Simulating
      two-dimensional fermionic condensed matter systems with matrix-product
      states.
    \newblock {\em Phys. Rev. B}, 104:075137, Aug 2021.
    
    \bibitem{Mate-2022}
    Mihály Máté, Klára Petrov, Szilárd Szalay, and {\"O}rs Legeza.
    \newblock Compressing multireference character of wave functions via fermionic
      mode optimization.
    \newblock {\em Journal of Mathematical Chemistry}, 61:362--375, 2023.
    
    \bibitem{Petrov-2023}
    Kl\'ara Petrov, \'Adam Ganyecz, Zsolt Benedek, Andr\'as Olasz, Gergely Barcza,
      and {\"O}rs Legeza.
    \newblock Low-cost generation of optimal molecular orbitals for multireference
      ci expansion: natural orbitals versus r´enyi entropy minimized orbitals
      provided by the density matrix renormalization group.
    \newblock In I.~Grabowski, K.~S\l{}owik, J.~Maruani, and E.~J. Br{\"a}ndas,
      editors, {\em Advances in Methods and Applications of Quantum Systems in
      Chemistry, Physics, and Biology Selected Proceed- ings of QSCP-XXV Conference
      (Torun, Poland, June 2022)}. Springer, Cham, 2024.
    
    \bibitem{Werner-2025}
    Miklós~Antal Werner, Andor Menczer, and {\"O}rs Legeza.
    \newblock Tensor network state methods and quantum information theory for
      strongly correlated molecular systems.
    \newblock {\em arXiv:2501.18263}, 2025.
    
    \bibitem{Friesecke-2023}
    Gero Friesecke, Gergely Barcza, and {\"O}rs Legeza.
    \newblock Predicting the {FCI} energy of large systems to chemical accuracy
      from restricted active space density matrix renormalization group
      calculations.
    \newblock {\em Journal of Chemical Theory and Computation}, 20(1):87--102,
      2023.
    
    \bibitem{Menczer-2024b}
    Andor Menczer, Maarten van Damme, Alan Rask, Lee Huntington, Jeff Hammond,
      Sotiris~S Xantheas, Martin Ganahl, and {\"O}rs Legeza.
    \newblock Parallel implementation of the density matrix renormalization group
      method achieving a quarter peta{FLOPS} performance on a single {DGX-H100 GPU}
      node.
    \newblock {\em Journal of Chemical Theory and Computation}, 20(19):8397--8404,
      2024.
    
    \bibitem{Menczer-2024e}
    Andor Menczer and {\"O}rs Legeza.
    \newblock Efficient calculation of one- and two-particle reduced density
      matrices on {GPU}s via the density matrix renormalization group method,
      unpublished, 2024.
    
    \bibitem{Legeza-2003c}
    {\"O}.~Legeza, J.~R\"oder, and B.~A. Hess.
    \newblock {QC-DMRG} study of the ionic-neutral curve crossing of {LiF}.
    \newblock {\em Molecular Physics}, 101(13):2019--2028, 2003.
    
    \bibitem{Barcza-2011}
    G.~Barcza, {\"O}.~Legeza, K.~H. Marti, and M.~Reiher.
    \newblock Quantum-information analysis of electronic states of different
      molecular structures.
    \newblock {\em Phys. Rev. A}, 83:012508, Jan 2011.
    
    \bibitem{Hammond-2007}
    Jeff~R. Hammond, Karol Kowalski, and Wibe~A. deJong.
    \newblock Dynamic polarizabilities of polyaromatic hydrocarbons using
      coupled-cluster linear response theory.
    \newblock {\em The Journal of Chemical Physics}, 127(14), October 2007.
    
    \bibitem{Hajgato-2009}
    B.~Hajgató, D.~Szieberth, P.~Geerlings, F.~De~Proft, and M.~S. Deleuze.
    \newblock A benchmark theoretical study of the electronic ground state and of
      the singlet-triplet split of benzene and linear acenes.
    \newblock {\em The Journal of Chemical Physics}, 131(22):224321, 12 2009.
    
    \bibitem{Zade-2010}
    Sanjio~S. Zade and Michael Bendikov.
    \newblock Heptacene and beyond: The longest characterized acenes.
    \newblock {\em Angewandte Chemie International Edition}, 49(24):4012--4015,
      2010.
    
    \bibitem{Pelzer-2011}
    Kenley Pelzer, Loren Greenman, Gergely Gidofalvi, and David~A. Mazziotti.
    \newblock Strong correlation in acene sheets from the active-space variational
      two-electron reduced density matrix method: Effects of symmetry and size.
    \newblock {\em The Journal of Physical Chemistry A}, 115(22):5632--5640, 2011.
    \newblock PMID: 21563790.
    
    \bibitem{Aiga-2012}
    Fumihiko Aiga.
    \newblock Theoretical study on oligoacenes and polycyclic aromatic hydrocarbons
      using the restricted active space self-consistent field method.
    \newblock {\em The Journal of Physical Chemistry A}, 116(1):663--669, 2012.
    \newblock PMID: 22201478.
    
    \bibitem{Rivero-2013}
    Pablo Rivero, Carlos~A. Jim{\'e}nez-Hoyos, and Gustavo~E. Scuseria.
    \newblock Entanglement and polyradical character of polycyclic aromatic
      hydrocarbons predicted by projected {H}artree–{F}ock theory.
    \newblock {\em The Journal of Physical Chemistry B}, 117(42):12750--12758,
      2013.
    \newblock PMID: 23668255.
    
    \bibitem{Plasser-2013}
    Felix Plasser, Hasan Pašalić, Martin~H. Gerzabek, Florian Libisch, Rafael
      Reiter, Joachim Burgdörfer, Thomas Müller, Ron Shepard, and Hans Lischka.
    \newblock The multiradical character of one- and two-dimensional graphene
      nanoribbons.
    \newblock {\em Angewandte Chemie International Edition}, 52(9):2581--2584,
      2013.
    
    \bibitem{Becke-1992}
    Axel~D Becke.
    \newblock Density-functional thermochemistry. {I}. the effect of the
      exchange-only gradient correction.
    \newblock {\em The Journal of chemical physics}, 96(3):2155--2160, 1992.
    
    \bibitem{Becke-1993}
    Axel~D Becke.
    \newblock A new mixing of {H}artree--{F}ock and local density-functional
      theories.
    \newblock {\em The Journal of chemical physics}, 98(2):1372--1377, 1993.
    
    \bibitem{Dunning-1989}
    Thom~H Dunning~Jr.
    \newblock Gaussian basis sets for use in correlated molecular calculations. i.
      the atoms boron through neon and hydrogen.
    \newblock {\em The Journal of chemical physics}, 90(2):1007--1023, 1989.
    
    \bibitem{Woon-1993}
    David~E Woon and Thom~H Dunning~Jr.
    \newblock Gaussian basis sets for use in correlated molecular calculations.
      {III}. the atoms aluminum through argon.
    \newblock {\em The Journal of chemical physics}, 98(2):1358--1371, 1993.
    
    \bibitem{Lippard-1994}
    Stephen~J Lippard and Jeremy~Mark Berg.
    \newblock {\em Principles of bioinorganic chemistry}.
    \newblock University Science Books, 1994.
    
    \bibitem{Izsak-2023}
    R{\'o}bert Izs{\'a}k, Aleksei~V. Ivanov, Nick~S. Blunt, Nicole Holzmann, and
      Frank Neese.
    \newblock Measuring electron correlation: The impact of symmetry and orbital
      transformations.
    \newblock {\em Journal of Chemical Theory and Computation}, 19(10):2703--2720,
      2023.
    \newblock PMID: 37022051.
    
    \bibitem{Faulstich-2019a}
    Fabian~M. Faulstich, Andre Laestadius, {\"O}rs Legeza, Reinhold Schneider, and
      Simen Kvaal.
    \newblock Analysis of the tailored coupled-cluster method in quantum chemistry.
    \newblock {\em SIAM Journal on Numerical Analysis}, 57(6):2579--2607, 2019.
    
    \bibitem{Liao-2024}
    Ke~Liao, Lexin Ding, and Christian Schilling.
    \newblock Quantum information orbitals ({QIO}): Unveiling intrinsic many-body
      complexity by compressing single-body triviality.
    \newblock {\em The Journal of Physical Chemistry Letters}, 15(26):6782–6790,
      June 2024.
    
    \bibitem{Larsson-2022}
    Henrik~R. Larsson, Huanchen Zhai, Klaas Gunst, and Garnet Kin-Lic Chan.
    \newblock Matrix product states with large sites.
    \newblock {\em Journal of Chemical Theory and Computation}, 18(2):749--762,
      2022.
    \newblock PMID: 35060382.
    
    \bibitem{Stolarczyk-1994}
    Leszek~Z. Stolarczyk.
    \newblock Complete active space coupled-cluster method. extension of
      single-reference coupled-cluster method using the {CASSCF} wavefunction.
    \newblock {\em Chemical Physics Letters}, 217(1–2):1 -- 6, 1994.
    
    \bibitem{Kurashige-2011}
    Yuki Kurashige and Takeshi Yanai.
    \newblock Second-order perturbation theory with a density matrix
      renormalization group self-consistent field reference function: Theory and
      application to the study of chromium dimer.
    \newblock {\em The Journal of Chemical Physics}, 135(9):094104, 2011.
    
    \bibitem{Pulay-2011}
    Peter Pulay.
    \newblock A perspective on the {CASPT2} method.
    \newblock {\em International Journal of Quantum Chemistry}, 111(13):3273--3279,
      2011.
    
    \bibitem{Sharma-2014b}
    Sandeep Sharma and Garnet Kin-Lic Chan.
    \newblock Communication: A flexible multi-reference perturbation theory by
      minimizing the {H}ylleraas functional with matrix product states.
    \newblock {\em The Journal of Chemical Physics}, 141(11):111101, 2014.
    
    \bibitem{Boguslawski-2011}
    Katharina Boguslawski, Konrad~H. Marti, and Markus Reiher.
    \newblock Construction of {CASCI}-type wave functions for very large active
      spaces.
    \newblock {\em The Journal of Chemical Physics}, 134(22):224101, 2011.
    
    \bibitem{Leyser-2024}
    Tiago Leyser~da Costa~Gouveia, Dimitrios Maganas, and Frank Neese.
    \newblock General spin-restricted open-shell configuration interaction
      approach: Application to metal {K}-edge {X}-ray absorption spectra of ferro-
      and antiferromagnetically coupled dimers.
    \newblock {\em The Journal of Physical Chemistry A}, 129(1):330–345, December
      2024.
    
    \bibitem{Legeza-2003b}
    {\"O}.~Legeza and J.~S\'olyom.
    \newblock Optimizing the density-matrix renormalization group method using
      quantum information entropy.
    \newblock {\em Phys. Rev. B}, 68:195116, Nov 2003.
    
    \end{thebibliography}
%

\end{document}